\documentclass[journal]{IEEEtran}
\usepackage{amsmath,amsfonts}
\usepackage{amsthm}
\usepackage[linesnumbered,ruled,vlined]{algorithm2e}
\usepackage{algorithmic}
\usepackage{array}
\usepackage{textcomp}
\usepackage{url}
\usepackage{verbatim}
\usepackage{graphicx}
\usepackage{caption}
\usepackage{placeins} % Add this in the preamble

\usepackage{bbm}
\usepackage{balance}
\usepackage[colorlinks]{hyperref}
\usepackage[dvipsnames]{xcolor}
\hypersetup{
    colorlinks = true,
    linkcolor  = red,       % color of internal links (e.g., equations, figures)
    citecolor  = blue,       % color of citations
    urlcolor   = blue        % color of external hyperlinks
}
\usepackage{enumitem}
\usepackage{stfloats}  % for widetext
\begin{document}
\title{Resource Allocation and AoI-Aware Detection for ISAC with Stacked Intelligent Metasurfaces}
 \author{\IEEEauthorblockN{Elaheh Ataeebojd,~\IEEEmembership{Member,~IEEE}, Nhan Thanh Nguyen,~\IEEEmembership{Senior Member,~IEEE}, Seonghoon Yoo, ~\IEEEmembership{Graduate Student Member,~IEEE}, Joonhyuk Kang, ~\IEEEmembership{Member,~IEEE}, Markku Juntti, \IEEEmembership{Fellow,~IEEE}, Matti Latva-aho, \IEEEmembership{Fellow,~IEEE}, and Mehdi Rasti, \IEEEmembership{Senior Member,~IEEE} %\\
        \thanks{This paper is accepted in part at the European Conference on Networks and Communications  (EuCNC), 2026 \cite{EuCNC2026}.}
 		\thanks{E. Ataeebojd, N. T. Nguyen, M. Juntti, M. Latva-aho, and  M. Rasti are with the Centre of Wireless Communications, University of Oulu, Oulu, Finland (email: \{elaheh.ataeebojd, nhan.nguyen, markku.juntti, matti.latva-aho, mehdi.rasti\}@oulu.fi).}
        \thanks{S. Yoo and J. Kang are with the School of Electrical Engineering, Korea Advanced Institute of Science and Technology, Daejeon 34141, South Korea (email: shyoo902@kaist.ac.kr, jhkang@ee.kaist.ac.kr).}
 		}
   }

\maketitle

\vspace{-2 em}
\begin{abstract} 
Stacked intelligent metasurfaces (SIMs) provide wave-domain degrees of freedom that can empower integrated sensing and communication (ISAC) through flexible beampattern synthesis and interference management, while reducing hardware cost. In this paper, we investigate energy-efficient resource allocation for a downlink SIM-aided multi-user ISAC system that supports the coexistence of enhanced mobile broadband (eMBB) and ultra-reliable and low-latency communication (URLLC) via puncturing, while simultaneously illuminating sensing targets. We formulate an energy efficiency (EE) maximization problem that jointly optimizes resource block (RB) allocation, transmit power control, and SIM phase shifts. The formulated problem is highly challenging due to the large number of variables optimized on different time scales. To overcome this, we leverage the intrinsic two-timescale structure induced by the puncturing approach to decompose the original problem into two tractable subproblems: EE maximization for eMBB users in each time slot and EE maximization for URLLC users and sensing targets in each mini-slot. To address each subproblem, we develop an iterative algorithm that transforms the original non-convex formulation into a sequence of tractable subproblems, yielding convex updates for RB allocation and power control, along with low-complexity updates for SIM phase shifts. Simulation results show that the proposed design achieves up to 230\% improvement in EE over a No-SIM baseline. In addition, it requires significantly fewer transmit antennas than conventional BS architectures, while preserving the EE achieved and satisfying the communication and sensing quality of service (QoS) requirements. Moreover, the results reveal fundamental trade-offs between EE and heterogeneous QoS requirements across communication and sensing functionalities.
\end{abstract}


\vspace{-1 em}
\section{Introduction}
Integrated sensing and communication (ISAC) has emerged as a core capability of the sixth-generation (6G) radio access networks (RANs), allowing simultaneous user connectivity and environment awareness through shared spectrum, hardware, and signal processing resources \cite{Survey_ISAC, ITU_20230, Liu_2022_JSAC}. This tight integration is motivated by emerging applications, such as autonomous systems, industrial automation, and large-scale Internet of Things (IoT) \cite{ISAC_book}, which require the joint support of communication, sensing, and control under stringent data rate, latency, reliability, and energy constraints \cite{ITU_20230, Liu_2022_JSAC}. As a result, multi-user ISAC systems must simultaneously satisfy heterogeneous communication requirements and sensing objectives, such as beampattern shaping and timely target detection. 

Since communication and sensing share the same physical-layer transmission, they compete for limited radio resources, including spectrum, transmit power, and spatial degrees of freedom. This coupling imposes fundamental challenges on scheduling and precoding, as the transmit signal must simultaneously support multiuser communication and sensing illumination requirements. To address this limitation, stacked intelligent metasurfaces (SIMs) have recently emerged as a promising technology for enabling wave-domain electromagnetic (EM) control \cite{SIM_2025, Wang_2024}. By cascading multiple programmable metasurface layers, SIMs provide enhanced flexibility compared to reconfigurable intelligent surfaces (RISs) \cite{Rasi_2024}. Moreover, by shifting part of the signal processing from the digital domain to the wave domain, SIM-assisted architectures can reduce hardware complexity and energy consumption compared with conventional base stations (BSs) equipped with fully digital antenna arrays \cite{An_2023, Abbas_2025, An_2024}.

Beyond the sensing--communication coupling, 6G networks are envisioned to support heterogeneous services, particularly enhanced mobile broadband (eMBB) and ultra-reliable and low-latency communication (URLLC) \cite{ITU_20230}. Specifically, eMBB aims to deliver extremely high data rates, while URLLC is designed for ultra-reliable and low-latency communications \cite{Rasti2022}. Although these service classes already exist in 5G, their stricter and inherently conflicting quality of service (QoS) requirements in 6G  make their joint support more challenging. To enable the coexistence of eMBB and URLLC, 3GPP \cite{TS38211} adopts a two-timescale scheduling framework, where resource blocks (RBs) are allocated to eMBB users over time slots (e.g., 1 ms) and remain fixed within each time slot. In contrast, URLLC traffic is sporadic and requires immediate service, and is therefore scheduled over shorter mini-slots (e.g., 0.125 ms) within a time slot. To efficiently multiplex these services, 3GPP recommended the puncturing (preemption) method within this two-timescale framework \cite{3gpp_700022, 3GPP_700204, Alsenwi2021, Prathyusha2022}. In the puncturing approach, a portion of the RBs allocated to ongoing eMBB transmissions is temporarily reallocated to URLLC. While puncturing enables higher access to resources for URLLC and improves spectrum utilization, it disrupts eMBB transmissions and reduces their achievable data rates. Consequently, resource allocation must carefully balance the conflicting requirements of eMBB and URLLC, a challenge that becomes more pronounced in ISAC systems where sensing constraints further limit transmit power and spatial degrees of freedom.

Sensing performance in ISAC critically depends on the timeliness of detection updates, which must be scheduled alongside URLLC arrivals and ongoing eMBB transmissions, making sensing timeliness a key constraint. Each target must be detected regularly; otherwise, sensing information becomes stale, and the BS may act on outdated states. However, frequent detections are resource-intensive and directly compete with communication resources--particularly under mini-slot operation, where URLLC traffic already consumes urgent resources. To capture the trade-off between detection timeliness and resource utilization, we adopt the age of information (AoI) metric, defined as the time elapsed since the most recent detection update. By constraining AoI, each target is guaranteed to be re-detected within a prescribed interval, ensuring timely sensing updates while complementing QoS requirements in ISAC.

Motivated by (i) wave-based control enabled by SIMs, (ii) the need to support heterogeneous services, (iii) AoI-aware detection timeliness, and (iv) the importance of energy efficiency (EE) in dense 6G deployments with periodic sensing, this paper investigates energy-efficient resource allocation for SIM-aided multi-user ISAC under heterogeneous QoS requirements for communication and sensing.

\vspace{-1 em}
\subsection{Related Works}
Existing works on SIM-aided wireless systems can be broadly categorized into \emph{communication-centric} SIM designs and, more recently, \emph{SIM-enabled ISAC}. For communication-centric designs, SIMs have been extensively studied to enhance sum data rate in multiple-input multiple-output (MIMO) networks \cite{An_2023, Papazafeiropoulos_2024, Papazafeiropoulos_2024_letter_CSI, Li_2024, Papazafeiropoulos_2024_letter_nearfield, Li_2025, Papazafeiropoulos_2025, Hu_2025, Shi_2025_June, An_2025, Huai_2025, Zhang_2025}. Specifically, in \cite{An_2023, Papazafeiropoulos_2024, Papazafeiropoulos_2024_letter_CSI, Li_2024}, SIM-assisted holographic MIMO transceivers were considered, where wave-domain precoding/combining is implemented via SIM phase shifts at the transmitter and/or receiver. For instance, in \cite{An_2023}, SIM phase shifts were optimized to shape the end-to-end channel, while in \cite{Papazafeiropoulos_2024}, SIM phase shifts and the transmit covariance matrix were jointly optimized. Under statistical channel state information (CSI), a joint optimization of SIM phase shifts and transmit power was addressed in \cite{Papazafeiropoulos_2024_letter_CSI}. Moreover, in \cite{Papazafeiropoulos_2024_letter_nearfield} and \cite{Li_2025}, the downlink direction of multiuser communications within the near field region was considered. In \cite{Papazafeiropoulos_2024_letter_nearfield}, the authors focused on the transmit power and SIM phase shifts optimization, whereas the joint design of the SIM phase shifts and digital precoding at the BS, accounting for SIM phase tuning errors, was tackled in \cite{Li_2025}. 
The authors of \cite{Papazafeiropoulos_2025} proposed deploying SIMs both at the BS side and in the intermediate space between the BS and users, further to shape the propagation environment in a massive MIMO network; the phase shifts of both SIMs were then jointly optimized to enhance the uplink data rate. The integration of SIMs into cell-free MIMO systems was investigated in \cite{Hu_2025, Shi_2025_June, Li_2024, Shi_2025_May}. The authors of \cite{Hu_2025} and \cite{Shi_2025_June} focused on downlink transmission, whereas uplink transmission was considered in \cite{Li_2024} and \cite{Shi_2025_May}. Specifically, in \cite{Hu_2025}, the transmit power in BSs and the phase shifts of the SIM were jointly optimized, while a joint optimization of user assignment, BS precoding, and SIM configurations was addressed in \cite{Shi_2025_June}. In \cite{Li_2024}, transmit/receive beamforming was coordinated with SIM phase shifts via alternating optimization. Moreover, under statistical CSI, the authors of \cite{Shi_2025_May} designed a gradient descent algorithm to optimize SIM phase shifts and a max-min power control algorithm to minimize the number of required BSs and antennas. Under a transmit power budget at the BS and discrete SIM phase shift constraints, multiuser downlink beamforming in the wave domain was investigated in \cite{An_2025} and \cite{Perovic_2025}, where the authors of \cite{An_2025} and \cite{Perovic_2025} focused on the sum data rate maximization and EE maximization, respectively. Although the above works demonstrate the potential of SIMs for communication-centric performance enhancement, most of them primarily focus on maximizing the sum data rate and do not address heterogeneous QoS requirements for users.

Communication-centric SIM designs have begun to move beyond purely the sum data rate maximization by incorporating QoS requirements and practical considerations in \cite{Huai_2025} and \cite{Zhang_2025}. In particular, in \cite{Huai_2025}, SIM-based wave-domain processing was integrated with the rate-splitting multiple access strategy to maximize the sum data rate under minimum data rate requirements for users. In addition, reliability in short-packet communications was taken into account using the finite blocklength (FBL) formulation in \cite{Zhang_2025}, which jointly optimized users' transmit power, SIM phase shifts, and receiver beamforming at the BS to maximize the sum data rate. Although these works confirm that SIM can support multiuser transmission under QoS requirements, energy-efficient SIM-based frameworks for the coexistence of heterogeneous services, such as eMBB with minimum data rate requirements and URLLC with stringent latency and reliability constraints, remain insufficiently addressed.

The integration of SIMs into ISAC has recently attracted attention in \cite{Niu_2024, Li_2025_ISAC, Wang_2024, Ranasinghe_2025, Ebrahimi_2025}, where SIMs provide wave-domain degrees of freedom to jointly support communication and sensing via beampattern shaping and/or estimation-driven designs. In \cite{Niu_2024}, the SIM was configured to synthesize a desired beampattern while serving multiple downlink communication users and detecting a sensing target. Consequently, SIM phase shifts and the transmit power of the BS were jointly optimized to maximize the sum data rate of users under beampattern constraints and a total transmit power budget. In \cite{Li_2025_ISAC}, a multi-objective problem was formulated to maximize the sum data rate of users while optimally shaping the normalized sensing beampattern for target detection, and a gradient ascent algorithm was proposed in \cite{Li_2025_ISAC} to solve the problem. The authors of \cite{Wang_2024} jointly optimized the transmit beamforming of the BS and SIM phase shifts by minimizing the Cram\'er--Rao bound (CRB) for target estimation subject to minimum signal-to-interference-plus-noise ratio (SINR) requirements for communication users and a transmit power budget at the BS, and further validated the approach via prototyping/experiments. In \cite{Ranasinghe_2025}, SIM phase shifts were optimized to maximize the effective channel gain of the weakest SIM--target--SIM cascaded path using a gradient ascent algorithm. SIM-enabled ISAC has also been extended to emerging scenarios, e.g., terahertz communications with an additional environmental RIS \cite{Ebrahimi_2025}. Despite these advances, existing SIM-enabled ISAC studies typically focus on data-rate-centric metrics (e.g., SINR or data rate) with beampattern/CRB-based sensing criteria; however, a service-aware QoS formulation that jointly accommodates heterogeneous communication requirements while ensuring sensing performance---together with energy-efficient resource allocation---remains underexplored.

\vspace{-1 em}
\subsection{Contributions}
Existing SIM-aided communication designs largely focus on maximizing the sum data rate \cite{An_2023, Papazafeiropoulos_2024, Papazafeiropoulos_2024_letter_CSI, Li_2024, Papazafeiropoulos_2024_letter_nearfield, Li_2025, Papazafeiropoulos_2025, Hu_2025, Shi_2025_June, An_2025, Huai_2025, Zhang_2025, Shi_2025_May}. Some related studies extend beyond maximizing the sum data rate. For instance, in \cite{Perovic_2025}, the authors formulated EE maximization. Also, minimum data rate requirements for eMBB users in \cite{Huai_2025} and reliability for URLLC users in \cite{Zhang_2025} were considered. However, these studies do not address the joint support of heterogeneous services. For example, they do not support both eMBB and URLLC at once, nor do they account for timeliness-driven sensing constraints. 
Furthermore, SIM-enabled ISAC designs typically optimize communication-centric metrics and sensing criteria such as beampattern gain or CRB \cite{Niu_2024, Li_2025_ISAC, Wang_2024, Ranasinghe_2025, Ebrahimi_2025}. Despite this, energy-efficient operation remains largely unexplored in these works.

In 6G and IoT systems with dense deployments, continuous sensing, and heterogeneous service requirements, EE becomes an important design objective \cite{ISAC_book}. These characteristics impose stringent power constraints on network operation, making energy-efficient ISAC design particularly critical. SIM architectures can significantly improve EE by enabling wave-domain control with reduced hardware overhead. Motivated by this, we investigate EE-oriented resource allocation for SIM-aided multi-user ISAC with heterogeneous communication and sensing requirements. Our proposed framework leverages a 3GPP-aligned two-timescale model to support the coexistence of eMBB and URLLC via puncturing, while ensuring sensing timeliness through AoI constraints and reliable target detection via directional beampattern control. To the best of our knowledge, this is the first work to jointly address these challenges within a unified EE-centric design. The major contributions of our paper are as follows:
\begin{itemize} [leftmargin=*]
    \item We develop a downlink SIM-aided multi-user ISAC architecture that jointly supports heterogeneous communication services and sensing under shared-waveform coupling and limited radio resources. Leveraging the 3GPP two-timescale structure and puncturing, we schedule eMBB on time slots and handle URLLC traffic and sensing updates on mini-slots via RB preemption.
    \item The joint RB allocation, transmit power control, and SIM phase shifts problem is optimized under both communication and sensing constraints. Communication requirements include minimum data rates for eMBB users as well as latency and reliability guarantees for URLLC users. Sensing constraints include beampattern gain requirements and AoI-based detection timeliness. To efficiently handle the resulting non-convex problem, we exploit the intrinsic two-timescale structure enabled by puncturing in 3GPP, where eMBB users are scheduled at each time slot, while URLLC users and sensing targets are handled at each mini-slot, to achieve a time-scale-aware decomposition of the joint optimization problem. Based on this decomposition, we develop an iterative algorithm using alternating optimization (AO) and fractional programming, which yields a sequence of tractable subproblems with convex updates for RB allocation and power control and low-complexity updates for SIM phase shifts.
    
    \item Simulation results demonstrate significant EE gains over baseline schemes.   Specifically, the proposed SIM architecture achieves up to 230\% improvement in EE over the No-SIM scheme. Furthermore, it matches or surpasses the performance of the conventional BS architecture with only 4 transmit antennas, whereas the conventional BS requires 24 or more for similar EE. Finally, the results reveal key trade-offs between EE and heterogeneous QoS requirements for communication and sensing.
\end{itemize}

\vspace{-1 em}
\subsection{Paper Organization and Notations}
The rest of this paper is organized as follows. Section~\ref{Sys} describes the system model, detailing the communication and sensing models. In Sections \ref{problem} and \ref{Solution}, we formally state the EE optimization problem and describe the proposed solution, respectively. In Section \ref{results}, simulation results are presented, and the paper is concluded in Section \ref{Con}.

Throughout the paper, scalars, vectors, and matrices are denoted by lower-case, boldface lower-case, and boldface uppercase letters, respectively. The operators $(\cdot)^T$ and $(\cdot)^H$ denote the transpose and Hermitian transpose, respectively. The notation $\mathcal{CN}(0, \sigma^2)$ represents the Gaussian distribution with mean $0$ and variance $\sigma^2$. The operator $\mathrm{diag}(\cdot)$ indicates a diagonal matrix, $\mathbb{E}[.]$ denotes the expectation, and $|.|$ refers to the magnitude of a complex number.

\section{System Model}\label{Sys} 
\begin{figure}[t!]	
    \centering
    \includegraphics[width=8 cm, height=3.5 cm]{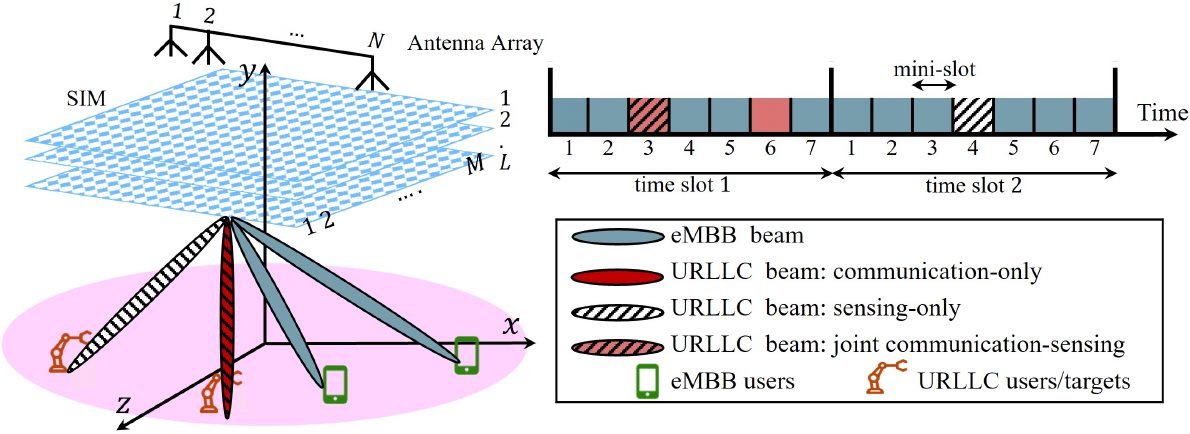}
    \vspace{-5 pt}
    \caption{Illustration of the proposed SIM-enabled ISAC with two-timescale scheduling. \label{fig:Sys}}
\end{figure}

We consider the downlink of a SIM-aided multicarrier MU-MIMO system employing orthogonal frequency-division multiple access (OFDMA), where the BS is equipped with $N$ antennas and employs a SIM. The BS serves two types of single-antenna users: eMBB users and URLLC communication users. Here, eMBB users are characterized by continuous traffic and high data rate requirements, whereas URLLC users exhibit sporadic arrivals with stringent latency and reliability constraints. Let $\mathcal{U}^{\rm{e}}$ and $\mathcal{U}^{\rm{u}}$ denote the sets of eMBB users and URLLC users, respectively, then the set of all users can be given as $\mathcal{U} = \mathcal{U}^{\rm{e}}\cup\mathcal{U}^{\rm{u}}$. Motivated by URLLC applications such as industrial automation and process control, where devices exhibit sporadic, event-driven traffic patterns and reliability is tightly coupled with timely situational awareness (e.g., device monitoring, localization, and hazard detection), the BS is required not only to transmit URLLC packets but also to monitor the URLLC-associated devices \cite{TS22104}, thereby enabling joint communication and sensing functionalities. Consequently, we treat each URLLC user as a sensing target, and the set of sensing targets coincides with $\mathcal{U}^{\rm{u}}$. Hereafter, the terms \emph{URLLC users} and \emph{sensing targets} are used interchangeably. Fig.~\ref{fig:Sys} illustrates the considered SIM-enabled ISAC system with two-timescale scheduling.

Employing OFDMA, the total frequency bandwidth centered at carrier frequency $f_c$ is equally divided into $C$ orthogonal RBs, indexed by $\mathcal{C} = \{1, 2, ..., C\}$, each with bandwidth $B$. In the frequency domain, each RB consists of 12 sub-carriers with $15\text{kHz}$ sub-carrier spacing, resulting in a bandwidth of $B=180$ kHz. In the time domain, following the 3GPP NR frame structure \cite{3GPP_700204}, the timeline is divided into time slots indexed by $\mathcal{T}=\{1,2,\ldots,T\}$, where each time slot has duration $1$~ms and contains 14 OFDM symbols. An RB spans 12 subcarriers over one time slot. To support the coexistence of eMBB and URLLC services, each time slot is further divided into $I$ mini-slots indexed by $\mathcal{I}=\{1,2,\ldots,I\}$. A mini-slot spans a small number of OFDM symbols (e.g., 2/4/7 symbols), and its duration is determined accordingly \cite{TS38211, 3GPP_700204}. Based on this two-timescale scheduling structure, also proposed in \cite{Alsenwi2021, Prathyusha2022}, eMBB traffic is scheduled over time slots, while URLLC is scheduled over mini-slots to satisfy stringent latency and reliability requirements. We assume that the number of arrival packets per mini-slot for each URLLC user $i\in \mathcal{U}^{\rm{u}}$ follows a Poisson distribution with mean $\lambda_{i}$.

\vspace{-1.5 em}
\subsection{SIM Model} \label{SIM_Model}
The SIM integrated with the BS is composed of $L$ programmable metasurface layers, each comprising $M$ meta-atoms with $M \!\! \geq \!\! N$. Let $\mathcal{L}\!\!= \!\! \{1,\ldots,L\}$ and $\mathcal{M} \!\!= \!\!\{1,\ldots,M\}$ denote the sets of layers and meta-atoms, respectively. An intelligent controller dynamically adjusts the complex transmission coefficients of the meta-atoms to manipulate the EM waves across layers, thereby enabling wave-domain beamforming and spatial multiplexing. 
% In addition, the SIM enables each BS antenna to transmit a dedicated stream to a specific user. 
Let $\phi_{m}^{(\ell)} \!\!\!\!=\!\!\! {e}^{j\theta_{m}^{(\ell)}}$ denote the transmission coefficient imposed by meta-atom $m$ on layer $\ell$, where $\theta_{m}^{(\ell)}\!\!\! \in \!\!\! [0,2\pi)$ represents the corresponding phase shift. The transmission coefficient vector of layer $\ell$ and its diagonal matrix representation are denoted by $\boldsymbol{\phi}^{(\ell)} \!\!=\!\! [\phi_{1}^{(\ell)}, ...,\phi_{M}^{(\ell)}]^{T} \!\!\in \!\! \mathbb{C}^{M\times1}$ and $\boldsymbol{\Phi}^{(\ell)} \!\!=\!\! {\rm{diag}}(\boldsymbol{\phi}^{(\ell)}) \!\!\in \!\! \mathbb{C}^{M\times M}$, respectively. Without loss of generality, each SIM layer is modeled as a uniform planar array with $M_x$ and $M_z$ meta-atoms along the $x$- and $z$-axes, respectively, and adjacent meta-atoms spaced by $\lambda/2$ on both axes, yielding $M \!\!=\!\! M_x \!\times \! M_z$ \cite{Liu_UPA}. Here, $\lambda\!\!= \!\! c/f_{c}$ denotes the wavelength and $c$ is the speed of light.

For simplicity, we assume that all layers are parallel and equally spaced. Thus, the spacing between two adjacent layers is $d = D/L$, where $D$ denotes the total thickness of the SIM. Let $r_{m,\tilde{m}}^{(\ell)}$ denote the propagation distance from meta-atom $\tilde{m}$ on layer $(\ell-1)$ to meta-atom $m$ on layer $\ell$ for $\ell \in\mathcal{L}\setminus\{1\}$. According to the Rayleigh--Sommerfeld diffraction theory \cite{Lin2018}, the transmission coefficient from meta-atom $\Tilde{m}$ on layer $(\ell - 1)$ to meta-atom $m$ on layer $\ell\in\mathcal{L}\setminus\{1\}$ is
\begin{align} \label{theory}
    \psi_{m, \Tilde{m}}^{(\ell)} = \frac{A \cos{\chi_{m, \Tilde{m}}^{(\ell)}} }{r_{m, \Tilde{m}}^{(\ell)}} \Big(\frac{1}{2\pi r_{m, \Tilde{m}}^{(\ell)}} -j\frac{1}{\lambda}\Big) e^{j2\pi r_{m, \Tilde{m}}^{(\ell)} / \lambda},
\end{align}
where $A$ is the area of each meta-atom in the SIM and $\chi_{m, \Tilde{m}}^{(\ell)}$ represents the angle between the propagation direction and the normal direction of layer $(\ell - 1)$. Hence, the transmission coefficient matrix between layer $(\ell - 1)$ and $\ell$ is given by $\mathbf{\Psi}^{(\ell)} \!\!\!=\!\!\! [\psi_{m, \Tilde{m}}^{(\ell)}]_{m, \Tilde{m}} \!\!\in \!\! \mathbb{C}^{M\times M}, \forall \ell \!\!\in\!\! \mathcal{L}\setminus\{1\}$. Unlike $\boldsymbol{\Phi}^{(\ell)}$, the matrix  $\mathbf{\Psi}^{(\ell)}$ is generally full, capturing diffraction-based coupling across meta-atoms.
Additionally, we assume that the BS's antennas are arranged in a uniform linear array with half-wavelength spacing. The array center is aligned with the centers of all metasurfaces. Accordingly, the distance from antenna $n$ to meta-atom $m$ on the input layer of the SIM is denoted by $r_{m, n}^{(1)}$, and the transmission coefficient matrix from the BS antenna array to the input layer of the SIM is represented by $\mathbf{\Psi}^{(1)}\!\!\!=\!\!\![\psi_{m, n}^{(1)}]_{m, n} \!\!\!\in\!\!\!\mathbb{C}^{M\times N}$, where its elements are computed similarly to \eqref{theory} using $r_{m,n}^{(1)}$ and $\chi_{m, n}^{(1)}$. Consequently, the wave-based beamforming matrix of the SIM is given by
\begin{align}
    \mathbf{\Theta} = \mathbf{\Phi}^{(L)} \mathbf{\Psi}^{(L)} \dots \mathbf{\Phi}^{(2)} \mathbf{\Psi}^{(2)} \mathbf{\Phi}^{(1)} \mathbf{\Psi}^{(1)} \in \mathbb{C}^{M\times N}.
\end{align}

\vspace{-1 em}
\subsection{Communication Model} \label{Com_Model}
We consider multiuser downlink transmissions in a single cell OFDMA network to both URLLC and eMBB users over $C$ orthogonal RBs assisted by the SIM. Let $s_{i, c}^{\rm{u}}[t, \tau]\sim\mathcal{CN}(0,1)$ denote the information symbol for URLLC user $i\in\mathcal{U}^{\rm{u}}$ on RB $c\in\mathcal{C}$ at mini-slot $\tau\in\mathcal{I}$ of time slot $t\in\mathcal{T}$ (i.e., mini-slot $(t, \tau)$), where symbols are independent across users and RBs. Therefore, in the OFDMA network, the transmit signal to URLLC user $i$ on RB $c$ at the SIM output is
\begin{equation*}
\mathbf{x}_{i, c}^{\mathrm{u}}[t, \tau]
=\beta_{i,c}[t,\tau]\,
[\mathbf{\Theta}[t, \tau]]_{:, i}\,
\sqrt{p_{i, c}^{\mathrm{u}}[t, \tau]}\,
s_{i, c}^{\mathrm{u}}[t, \tau]
\in \mathbb{C}^{M \times 1},
\end{equation*}
where $p_{i, c}^{\rm{u}}[t, \tau]$ is the transmit power allocated to URLLC user $i$ on RB $c$ at $(t,\tau)$, $[\mathbf{\Theta}[t, \tau]]_{:, i}$ denotes the SIM-induced wave-domain beamforming vector for URLLC user $i$ at mini-slot $(t,\tau)$, which is determined by the SIM coefficients $\{\boldsymbol{\Phi}^{(\ell)}\}_{\ell=1}^{L}$, and $\beta_{i, c}[t, \tau]\in\{0,1\}$ denotes the RB allocation indicator for URLLC users at mini-slot $(t,\tau)$. If RB \(c\) is assigned to URLLC user $i$, then \(\beta_{i, c}[t, \tau]=1\); otherwise \(\beta_{i, c}[t, \tau]=0\). Due to OFDMA, we have $\sum_{i\in\mathcal{U}^{\rm{u}}} \beta_{i,c}[t,\tau]\le 1, \forall c\in\mathcal{C}, t\in\mathcal{T}, \tau\in\mathcal{I}$.

Due to the spatial correlation induced by the densely packed meta-atoms in the SIM, the channel from the SIM output layer to URLLC users is assumed to follow a spatially correlated Rayleigh fading model \cite{An_2024, An_2024_ISAC, Li_2024}.\footnote{This assumption is adopted for tractable modeling and performance evaluation; however, the proposed framework is applicable to general channel realizations.} Specifically, the channel vector from the SIM output to URLLC user $i$ on RB $c$, denoted by $\mathbf{h}_{i, c}^{\rm{u}}[t, \tau] \in \mathbb{C}^{M \times 1}$, is modeled as 
$\mathbf{h}_{i, c}^{\rm{u}}[t, \tau] \sim \mathcal{CN}\big(0, \upsilon_{i} \mathbf{R}_{\rm{SIM}}\big)$, where $\upsilon_{i}$ accounts for the large-scale path loss between the SIM and user $i$, modeled as $\upsilon_{i} = C_{0} (\frac{D_{i}}{D_{0}})^{-\alpha}$. Here, $D_{i}$ is the propagation distance to user $i$, $C_{0} = ( \frac{\lambda}{4\pi D_0})^2$ is the free-space path loss at the reference distance $D_0 = 1$m, and $\alpha$ is the path loss exponent. The matrix $\mathbf{R}_{\rm{SIM}} \in \mathbb{C}^{M \times M}$ denotes the spatial correlation among the meta-atoms of the SIM. Under isotropic scattering with uniformly distributed multipath components, $[\mathbf{R}_{\rm{SIM}}]_{m,\Tilde{m}} = \mathrm{sinc}( \frac{2r_{m,\Tilde{m}}}{\lambda})$ \cite{Emil_2021}, where $r_{m,\Tilde{m}}$ is the distance between meta-atom $\tilde{m}$ and meta-atom $m$ on the same layer. Therefore, the received signal at URLLC user $i$ on RB $c$ at $(t,\tau)$ is given by
\begin{equation}
    y_{i, c}^{\rm{u}}[t, \tau] = {\mathbf{h}_{i, c}^{{\rm{u}}\: H}[t, \tau]} \mathbf{x}_{i,c}^{\rm u}[t,\tau] + n_{i, c}^{\rm{u}}[t, \tau],
\end{equation}
where $n_{i, c}^{\rm{u}}[t, \tau] \sim \mathcal{CN}(0, \sigma_{i}^2)$ represents additive white Gaussian noise (AWGN) with variance $\sigma_{i}^2$. The received signal-to-noise-ratio (SNR) at URLLC user $i$ on RB $c$ at $(t,\tau)$ is
\begin{equation*}\label{SINR_URLLC}
\gamma_{i, c}^{\rm{u}}[t, \tau]
=({\beta_{i,c}[t,\tau]\big|\mathbf{h}_{i, c}^{{\rm{u}}\: H}[t, \tau] [\mathbf{\Theta}[t, \tau]]_{:, i} \big|^{2}p_{i, c}^{\rm{u}}[t, \tau]})\: / \:{\sigma_{i}^{2}}.
\end{equation*}
Due to the short blocklength and stringent reliability requirements of URLLC transmissions, Shannon's capacity formula, which assumes infinite blocklength, is not accurate. Therefore, the achievable data rate of URLLC user $i$ on RB $c$ is computed using the FBL capacity formula as \cite{Polyanskiy2010}
\begin{equation} 
\begin{aligned} \label{rate_URLLC} 
r_{i, c}^{\rm{u}}[t, \tau] =& \frac{B}{I}\big[ \log_2 ( 1 + \gamma_{i, c}^{\rm{u}}[t, \tau]) \!\!-\!\! \sqrt{\dfrac{V_{i, c}^{\rm{u}}[t, \tau]}{T_{b}}} Q^{-1}(\epsilon) \log_2 e\big], 
\end{aligned} 
\end{equation} 
where $Q^{-1}(.)$ is the inverse of Gaussian Q-function, $\epsilon$ denotes a desirable decoding error probability to assure the reliability of URLLC users, $T_{b}$ is the blocklength in symbols, and $V_{i, c}^{\rm{u}}[t, \tau]= 1- (1+\gamma_{i, c}^{\rm{u}}[t, \tau])^{-2}$ is the channel dispersion. When the received SNR exceeds 5 dB, $V_{i, c}^{\rm{u}}[t, \tau]$ can be accurately approximated as 1 in cellular networks. On the other hand, in a low SNR regime where $V_{i, c}^{\rm{u}}[t, \tau] < 1$, we can obtain a lower bound for the achievable data rate in (\ref{rate_URLLC}). By applying this lower bound to optimize resource allocation, we can further satisfy the latency and reliability requirements. Therefore, $r_{i, c}^{\rm{u}}[t, \tau]$ can be written as $r_{i, c}^{\rm{u}}[t, \tau] = B/I\left[ \log_2 \left( 1 + \gamma_{i, c}^{\rm{u}}[t, \tau] \right) - \sqrt{\frac{1}{T_{b}}}Q^{-1}(\epsilon) \log_{2} e\right]$. The total data rate of URLLC user $i$ is expressed as 
\begin{equation} 
\begin{aligned} 
r_{i}^{\rm{u}}[t, \tau] = \sum_{c\in \mathcal{C}} \beta_{i, c}[t, \tau] r_{i, c}^{\rm{u}}[t, \tau], \quad \forall i \in \mathcal{U}^{\rm{u}}.
\end{aligned} 
\end{equation} 
The end-to-end (E2E) delay consists of queueing delay, computation delay, propagation delay, and transmission delay. The transmission delay for sending a packet of size $L_{i}$ (in bits) is obtained by $T_{i}^{\text{trans}}[t, \tau] \!\!=\!\! L_{i} / r_{i}^{\rm{u}}[t, \tau]$. The propagation delay is negligible, while the computation delay is upper-bounded by a predefined threshold $T_{\mathrm{comp}}^{\mathrm{max}}$. Since transmission from the BS to URLLC users is single-hop \cite{Chen_2021}, and URLLC packets are transmitted within the upcoming mini-slot \cite{3gpp_700022}, the queueing delay is negligible compared to $T_{i}^{\mathrm{trans}}$ and $T_{\mathrm{comp}}^{\mathrm{max}}$. Therefore, the E2E delay is expressed as $T_{i, \text{e2e}}[t, \tau] = T_{i, \text{trans}}[t, \tau] + T_{\text{comp}}^{\text{max}}$. 

We denote the information symbol for eMBB user $i\in\mathcal{U}^{\rm{e}}$ on RB $c\in\mathcal{C}$ at time slot $t\in\mathcal{T}$ as $s_{i, c}^{\rm{e}}[t]\sim\mathcal{CN}(0,1)$, where symbols are independent across users and RBs. Therefore, the transmit signal to eMBB user $i$ on RB $c$ at the SIM output is
\begin{equation}
    \mathbf{x}_{i, c}^{\rm{e}}[t] = \alpha_{i,c}[t] [\mathbf{\Theta}[t]]_{:, i}\: \sqrt{p_{i, c}^{\rm{e}}[t]} s_{i, c}^{\rm{e}}[t] \in \mathbb{C}^{M \times 1}, 
\end{equation}
where $p_{i, c}^{\rm{e}}[t]$ is the transmit power allocated to eMBB user $i$ on RB $c$ at $t$, $[\mathbf{\Theta}[t]]_{:, i}$ denotes the SIM-induced wave-domain beamforming vector used to serve user $i$ at time slot $t$, which is determined by the SIM coefficients $\{\boldsymbol{\Phi}^{(\ell)}\}_{
\ell=1}^{L}$, and $\alpha_{i, c}[t]\in\{0,1\}$ denotes the RB allocation indicator for eMBB users at time slot $t$. If RB \(c\) is assigned to eMBB user $i$, then \(\alpha_{i, c}[t]=1\); otherwise \(\alpha_{i, c}[t]=0\). Due to OFDMA, we have $\sum_{i\in\mathcal{U}^{\rm{e}}} \alpha_{i,c}[t]\le 1,\: \forall c\in\mathcal{C},\; t\in\mathcal{T}$.

Likewise, we model the channel from the SIM output layer to eMBB users as a spatially correlated Rayleigh fading channel. Specifically, the channel vector from the SIM output to eMBB user $i$ on RB $c$, denoted by $\mathbf{h}_{i, c}^{\rm{e}}[t] \in \mathbb{C}^{M \times 1}$, is modeled as 
$\mathbf{h}_{i, c}^{\rm{e}}[t] \sim \mathcal{CN}\big(0, \upsilon_{i} \mathbf{R}_{\rm{SIM}}\big)$. Therefore, the received signal at eMBB user $i$ on RB $c$ at time slot $t$ is given by
\begin{equation}
    y_{i, c}^{\rm{e}}[t] = {\mathbf{h}_{i, c}^{{\rm{e}}\: H}[t]} \mathbf{x}_{i, c}^{\rm{e}}[t] + n_{i, c}^{\rm{e}}[t],
\end{equation}
where $n_{i, c}^{\rm{e}}[t] \sim \mathcal{CN}(0, \sigma_{i}^2)$ represents AWGN with variance $\sigma_{i}^2$. The SNR at eMBB user $i$ on RB $c$ at $t$ is
\begin{equation}\label{SINR_eMBB}
\gamma_{i, c}^{\rm{e}}[t]
=({\alpha_{i,c}[t]\big|\mathbf{h}_{i, c}^{{\rm{e}}\: H}[t] [\mathbf{\Theta}[t]]_{:, i} \big|^{2}p_{i, c}^{\rm{e}}[t]}) \: / \: {\sigma_{i}^{2}}.
\end{equation}
Owing to the relatively long blocklength of eMBB transmissions, Shannon's capacity formula accurately characterizes the achievable data rate. However, under the puncturing approach, eMBB users experience a loss in their achievable data rate. Consequently, the achievable data rate of eMBB user $i$ on RB $c$ is derived by
\begin{equation} 
\begin{aligned} \label{rate_eMBB} 
r_{i, c}^{\rm{e}}[t] =& \eta_{i, c}[t]\, B\, \log_2 \left( 1 + \gamma_{i, c}^{\rm{e}}[t] \right),
\end{aligned} 
\end{equation}
where
\[
\eta_{i, c}[t] = 1 - \frac{1}{I}\sum_{j\in \mathcal{U}^{\rm{u}}}\sum_{\tau\in \mathcal{I}} \alpha_{i, c}[t]\beta_{j, c}[t, \tau]
\]
captures the fraction of eMBB resources remaining after the puncturing of RBs by URLLC transmissions. The total data rate of eMBB user $i$ is expressed as 
\begin{equation} 
\begin{aligned} 
r_{i}^{\rm{e}}[t] = \sum_{c\in \mathcal{C}} \alpha_{i, c}[t] r_{i, c}^{\rm{e}}[t], \quad \forall i \in \mathcal{U}^{\rm{e}}.
\end{aligned} 
\end{equation}

\vspace{-2 em}
\subsection{Sensing Model} \label{Sensing_Model}
The BS detects URLLC users (targets) using the same resources that simultaneously support URLLC transmissions. The detectability of target $i\in \mathcal{U}^{\rm{u}}$ is characterized by the beampattern gain of the SIM. The beampattern gain directed toward target $i$ on RB $c$ is
\begin{equation*}
    P_{i, c}(\theta_{i}, \varphi_{i}; t, \tau) = 
    \mathbf{a}(\theta_{i}, \varphi_{i})^H \mathbf{\Sigma}_{c}[t, \tau]  \mathbf{a}(\theta_{i}, \varphi_{i}),
\end{equation*}
where $\mathbf{\Sigma}_{c}[t, \tau] \in\mathbb{C}^{M\times M}$ denotes the effective transmit covariance for RB $c$, obtained as
\begin{equation*}
    \begin{aligned}
        \mathbf{\Sigma}_{c}[t, \tau]
        &= \mathbb{E}\!\Big[
        \sum_{j\in\mathcal{U}^{\rm u}} \mathbf{x}_{j,c}^{\rm u}[t,\tau]
        \big(\sum_{k\in\mathcal{U}^{\rm u}} \mathbf{x}_{k,c}^{\rm u}[t,\tau]\big)^H
        \Big] \\
        &= \sum_{j \in \mathcal{U}^{\rm u}}
        \mathbb{E}\!\left[
        \mathbf{x}_{j,c}^{\rm u}[t,\tau]\mathbf{x}_{j,c}^{\rm u}[t,\tau]^H
        \right] \\
        &= \sum_{j \in \mathcal{U}^{\rm u}}
        p_{j, c}^{\rm u}[t, \tau]\,
        [\mathbf{\Theta}[t, \tau]]_{:, j}
        [\mathbf{\Theta}[t, \tau]]_{:, j}^H \, ,
    \end{aligned}
\end{equation*}
and $\mathbf{a}(\theta_{i}, \varphi_{i})$ is the steering vector of the SIM for target $i$ located at azimuth angle $\varphi_{i}$ and elevation angle $\theta_{i}$. Assuming that the channel between the SIM and the targets is modeled as a line-of-sight propagation channel, $\mathbf{a}(\theta_{i}, \varphi_{i})$ is defined as 
\begin{equation*}
    \mathbf{a}(\theta_{i}, \varphi_{i}) = \frac{1}{\sqrt{M_{x} M_{z}}} \, 
    \mathbf{a}_{x}(\theta_{i}, \varphi_{i}) \otimes \mathbf{a}_{z}(\theta_{i})\in \mathbb{C}^{M \times 1}, 
\end{equation*}
where $\mathbf{a}_x(\theta_{i}, \varphi_{i})\in \mathbb{C}^{M_x \times 1}$ and $\mathbf{a}_{z}(\theta_{i})\in \mathbb{C}^{M_z \times 1}$ are given by 
$\mathbf{a}_x(\theta_i,\phi_i)
=\big[1, e^{-j\pi \sin\theta_i \sin\phi_i}, \ldots, e^{-j\pi(M_x-1) \sin\theta_i \sin\phi_i}\big]^T$ and $\mathbf{a}_z(\theta_i)
=\big[1, e^{-j\pi \cos\theta_i}, \ldots, e^{-j\pi(M_z-1) \cos\theta_i}\big]^T$, respectively, with $\theta_{i} \in (0, \pi)$, and $\varphi_{i} \in (-\pi/2, \pi/2)$. 

Ensuring high instantaneous detectability in a given mini-slot is not sufficient, since each target must be revisited regularly; otherwise, the detection information becomes stale, and the BS may act on outdated target states. However, enforcing frequent detection updates for all targets is resource-consuming. Therefore, we define the AoI as the time elapsed since the most recent detection update, so bounding AoI is equivalent to ensuring that each target is detected within a prescribed update interval. We denote the AoI of target $i$ at mini-slot $(t, \tau)$ by $\delta_{i}[t, \tau]$. If target $i$ is detected at mini-slot $(t, \tau)$, then its AoI is reset to $1$ at the beginning of next mini-slot; otherwise, it increases by $1$. Accordingly, assuming that each target can be detected on at most one RB per mini-slot, the AoI evolves as
\begin{equation} \label{AOI_def}
    \delta_{i}[t, \tau+1] =
    \begin{cases}
        1, & \!\!\text{if } \sum_{c \in \mathcal{C}}\rho_{i, c}[t, \tau] = 1, \\
        \delta_{i}[t, \tau] + 1, & \!\!\!\text{otherwise,}
    \end{cases}
     \:\forall \tau \in \mathcal{I},
\end{equation}
where $\rho_{i, c}[t, \tau] \in \{0, 1\}$ denotes the RB allocation for target $i \in \mathcal{U}^{\rm{u}}$ at mini-slot $(t, \tau)$. If the BS detects target $i$ on RB $c$ at mini-slot $(t, \tau)$, then $\rho_{i, c}[t, \tau] = 1$; otherwise, $\rho_{i, c}[t, \tau] = 0$.

At the boundary between consecutive time slots, the AoI carries over according to
\begin{equation}
    \delta_{i}[t+1,1] = 
    \begin{cases}
        1, & \text{if } \sum_{c \in \mathcal{C}}\rho_{i, c}[t, I] = 1, \\
        \delta_{i}[t,I] + 1, & \text{otherwise.}
    \end{cases}
\end{equation}

For initialization, we set $\delta_{i}[1, 0] = 0$. To ensure that target detectability is maintained with new detection updates, we complement $P_{i,c}(\theta_i,\phi_i;t, \tau)$ with a constraint on the long-term average AoI for each target $i$, defined as
\begin{equation}
    \bar{\Delta}_{i} = 
    \lim_{T \to \infty} \lim_{I \to \infty}
    \frac{1}{T\times I} \sum_{t=1}^{T} \sum_{\tau=1}^{I}
    \mathbb{E}[\delta_{i}[t, \tau]].
\end{equation}

\vspace{-0.8 em}
\section{Problem Formulation} \label{problem}
In this section, we formally state the energy-efficient resource allocation problem for the proposed SIM-enabled joint communication and sensing framework. The scheduling decisions for eMBB users are made at the beginning of each time slot. Then, within a given time slot, the incoming traffic of URLLC users is served at each mini-slot by puncturing RBs previously assigned to eMBB transmissions. Meanwhile, the SIM phase responses are configured at every mini-slot to jointly support the communication links of both eMBB and URLLC users, while ensuring reliable detection of the targets in their corresponding angular directions. The instantaneous EE of the eMBB users at time slot $t$ is calculated by
\begin{equation}
    \begin{aligned}
        {EE}^{\rm{e}}[t] =\frac{\sum_{i\in\mathcal{U}^{\rm{e}}} r_{i}^{\rm{e}}[t]}{P_{\mathrm{tot}}^{\rm{e}}[t]},
    \end{aligned}
\end{equation}
where \(P_{\mathrm{tot}}^{\rm{e}}[t] = \sum_{i\in\mathcal{U}^{\rm{e}}}\sum_{c\in\mathcal{C}} \alpha_{i,c}[t] \, p_{i,c}^{\rm{e}}[t]\) denotes the total transmit power consumption of eMBB users at time slot \(t\). Similarly, the instantaneous EE of the URLLC users at mini-slot $(t, \tau)$ is given by
\begin{equation}
    \begin{aligned} \label{EE_URLLC}
    {EE}^{\rm{u}}[t, \tau] = \frac{\sum_{i\in\mathcal{U}^{\rm{u}}} r_{i}^{\rm{u}}[t, \tau]}{P_{\mathrm{tot}}^{\rm{u}}[t, \tau]},
    \end{aligned}
\end{equation}
where 
\begin{equation} \label{total_P_URLLC}
    \begin{aligned}
    P_{\mathrm{tot}}^{\rm{u}}[t, \tau] &= \sum_{i\in \mathcal{U}^{\rm{u}}} \sum_{c\in \mathcal{C}}\, (\beta_{i, c}[t, \tau] + \rho_{i, c}[t, \tau] \\
    & - \beta_{i, c}[t, \tau]\, \rho_{i, c}[t, \tau])\, p_{i, c}^{\rm{u}}[t, \tau].
    \end{aligned}
\end{equation}
As seen from \(P_{\mathrm{tot}}^{\rm u}[t,\tau]\), more frequent sensing updates, captured by $\rho_{i,c}[t,\tau]$, increase the transmit power consumption of URLLC users, thereby reducing their EE. Since sensing timeliness is enforced through AoI constraints, this introduces a trade-off between sensing freshness and EE.

To jointly evaluate communication and sensing performance, we adopt the time-averaged multi-service EE, defined over eMBB and URLLC transmissions, as the communication metric. For sensing, we use the beampattern gain to characterize the spatial focusing capability of the transmitted signal toward the targets, as it directly reflects the quality of directional target illumination in ISAC. Therefore, we aim at the joint optimization of RB allocation, transmit power allocation, and SIM phase shifts to maximize the time-averaged multi-service EE, while ensuring that eMBB users meet their minimum data rate requirements, URLLC users satisfy their delay and reliability constraints, and sensing requirements on beampattern gain and the long-term average AoI are fulfilled. This problem can be formulated as:
\begin{equation} \label{Problem_1}
\begin{aligned} 
     &\max_{\substack{
        \{\phi_{m}^{(\ell)}[t], \phi_{m}^{(\ell)}[t, \tau]\} \\
        \{\alpha_{i, c}[t], \beta_{i, c}[t, \tau], \rho_{i, c}[t, \tau]\} \\
        \{p_{i, c}^{\rm{e}}[t], p_{i, c}^{\rm{u}}[t, \tau]\} \\
    }} 
    \frac{1}{T} \sum_{t \in \mathcal{T}} {EE}^{\rm{e}}[t] 
    + \frac{1}{T} \sum_{t \in \mathcal{T}} \sum_{\tau \in \mathcal{I}} {EE}^{\rm{u}}[t, \tau] \\[0.5em]
    &\text{s.t.} \\
    & \text{C1: } r_{i}^{\rm{e}}[t]\geq r_{i}^{\rm{min}}, \quad \forall i \in \mathcal{U}^{\rm{e}}, \forall t \in \mathcal{T}, \\ 
    &\text{C2: } \mathbb{P}\left(r_{i}^{\rm{u}}[t, \tau]\! \geq \! {N}_{i}[t, \tau] L_{i} \right) \! \geq \! \gamma_{i}^{\rm{Rel}}, \forall i \in \mathcal{U}^{\rm{u}}, \forall \tau \in \mathcal{I}, \forall t \in \mathcal{T}, \\ 
    &\text{C3: } T_{i, \rm{e2e}}[t, \tau] \leq T_{i}^{\rm{max}}, \quad \forall i \in \mathcal{U}^{\rm{u}}, \forall \tau \in \mathcal{I}, \forall t \in \mathcal{T}, \\ 
     &\text{C4: } \bar{\Delta}_{i} \leq \Delta_{i}^{\rm{max}}, \quad \forall i \in \mathcal{U}^{\rm{u}}, \\
     &\text{C5: } \rho_{i, c}[t, \tau] \dfrac{P_{i, c}(\theta_{i}, \varphi_{i}; t, \tau)}{\upsilon_{i}^{2}} \geq \rho_{i, c}[t, \tau] \Gamma^{\rm{th}}, \\
     &\qquad\qquad\qquad\qquad\qquad \forall i \in \mathcal{U}^{\rm{u}}, \forall c \in \mathcal{C}, \forall \tau \in \mathcal{I}, \forall t \in \mathcal{T}, \\
     &\text{C6: } \sum_{i \in \mathcal{U}^{\rm{e}}} \alpha_{i, c}[t] \leq 1, \quad \forall c \in \mathcal{C}, \forall t \in \mathcal{T}, \\
    &\text{C7: } \sum_{i \in \mathcal{U}^{\rm{u}}} \beta_{i, c}[t, \tau] \leq 1, \quad \forall c \in \mathcal{C}, \forall \tau \in \mathcal{I}, \forall t \in \mathcal{T}, \\
      &\text{C8: } \sum_{i \in \mathcal{U}^{\rm{u}}} \rho_{i, c}[t, \tau] \leq 1, \quad \forall c \in \mathcal{C}, \forall \tau \in \mathcal{I}, \forall t \in \mathcal{T}, \\
     &\text{C9: } \sum_{c \in \mathcal{C}} \rho_{i, c}[t, \tau] \leq 1, \quad \forall i \in \mathcal{U}^{\rm{u}}, \forall \tau \in \mathcal{I}, \forall t \in \mathcal{T}, \\
     & \text{C10: } P_{\mathrm{tot}}^{\rm{e}}[t]+ P_{\mathrm{tot}}^{\rm{u}}[t, \tau] \leq P^{\rm{max}}, \quad \forall \tau \in \mathcal{I}, \forall t \in \mathcal{T}, \\
    & \text{C11: }|\phi_{m}^{(\ell)}[t]| = 1, \quad \forall m \in \mathcal{M}, \forall \ell\in \mathcal{L}, \forall t \in \mathcal{T}, \\ 
    & \text{C12: }|\phi_{m}^{(\ell)}[t, \tau]| = 1, \quad \forall m \in \mathcal{M}, \forall \ell\in \mathcal{L}, \forall \tau \in \mathcal{I}, \forall t \in \mathcal{T}, \\ 
    & \text{C13: }\alpha_{i, c}[t] \in \{0,1\}, \quad \forall i \in \mathcal{U}^{\rm{e}}, \forall c\in \mathcal{C}, \forall t \in \mathcal{T}, \\
    & \text{C14: }\beta_{i, c}[t, \tau] \in \{0,1\},
    \quad \forall i \in \mathcal{U}^{\rm{u}}, \forall c\in \mathcal{C}, \forall \tau \in \mathcal{I}, \forall t \in \mathcal{T}, \\
    & \text{C15: }\rho_{i, c}[t, \tau] \in \{0,1\},
    \quad \forall i \in \mathcal{U}^{\rm{u}}, \forall c\in \mathcal{C}, \forall \tau \in \mathcal{I}, \forall t \in \mathcal{T},\\
    & \text{C16: }p_{i, c}^{\rm{e}}[t] \geq 0,
    \quad \forall i \in \mathcal{U}^{\rm{e}}, \forall c\in \mathcal{C}, \forall t \in \mathcal{T},\\
    & \text{C17: }p_{i, c}^{\rm{u}}[t, \tau] \geq 0,
    \quad \forall i \in \mathcal{U}^{\rm{u}}, \forall c\in \mathcal{C}, \forall \tau \in \mathcal{I}, \forall t \in \mathcal{T},
\end{aligned}
\end{equation}
where constraint C1 ensures that each eMBB user $i$ meets its minimum data rate requirement, denoted by $r_{i}^{\rm{min}}$. In addition, C2 guarantees the reliability requirement for URLLC users by ensuring that the number of arrived packets $N_{i}[t, \tau]$ for each URLLC user $i$ at mini-slot $(t, \tau)$ is transmitted within mini-slot $(t, \tau)$ with probability of at least $\gamma_{i}^{\rm{Rel}}$. C3 enforces that the E2E delay for each data packet of URLLC user $i$ remains within the threshold $T_{i}^{\rm{max}}$. Furthermore, C4 limits the long-term average AoI of each target to its maximum tolerable threshold $\Delta_{i}^{\max}$. C5 ensures that the beampattern gain in the direction of the scheduled target meets the detection threshold $\Gamma^{\rm{th}}$, and $v_{i}^{2}$ captures the corresponding pathloss normalization. C6--C9 enforce exclusive RB assignment in the OFDMA system. Furthermore, C10 limits the total transmit power to the maximum BS transmit power $P^{\rm{max}}$. C11 describes the unit-modulus constraint on the SIM coefficients. C12--C14 represent the binary nature of the RB allocation and sensing selection variables. Finally, C15 and C16 represent the non-negative transmit power for eMBB and URLLC transmissions on each RB, respectively.

\begin{algorithm}[t!]
\SetAlCapFnt{\footnotesize}
\SetAlCapNameFnt{\footnotesize}
\caption{Overall proposed two-timescale algorithm} \label{alg:overall}
\footnotesize
\begin{algorithmic}[1]
\STATE \textbf{Input:} Initial AoI $\delta_i[1,0]=0$, virtual queues $U_i[1,0]=0$, $\forall i\in\mathcal{U}^{\rm u}$, and initial SIM phases $\phi_{m}^{(\ell)}[1,0]$.
\FOR{each time slot $t\in\mathcal{T}$}
    \STATE Solve problem \eqref{Prob_eMBB} using Algorithm~\ref{alg:eMBB} and obtain $\{\alpha_{i,c}^{\star}[t],p_{i,c}^{{\rm e},\star}[t], {\phi}_{m}^{(\ell), \star}[t]\}$.
    \FOR{each mini-slot $\tau\in\mathcal{I}$}
        \STATE Solve problem \eqref{Prob_URLLC} using Algorithm~\ref{alg:URLLC} and obtain $\{\beta_{i,c}^{\star}[t,\tau],\rho_{i,c}^{\star}[t,\tau],p_{i,c}^{{\rm u},\star}[t,\tau],\phi_{m}^{(\ell), \star}[t,\tau]\}$.
        \STATE Compute $\rho_i[t,\tau]=\sum_{c\in\mathcal{C}}\rho_{i,c}^{\star}[t,\tau]$, $\forall i\in\mathcal{U}^{\rm u}$.
        \STATE Update AoI $\delta_i[t,\tau+1]$ via \eqref{aoi_rec}, $\forall i\in\mathcal{U}^{\rm u}$.
        \STATE Update virtual queues $U_i[t,\tau+1]$ via \eqref{eq:virtual_queue_update}, $\forall i\in\mathcal{U}^{\rm u}$.
    \ENDFOR
\ENDFOR
\STATE \textbf{Output:} $\{\alpha_{i,c}^{\star}[t],p_{i,c}^{{\rm e},\star}[t], \phi_{m}^{(\ell), \star}[t]\}$ and $\{\beta_{i,c}^{\star}[t, \tau],\rho_{i,c}^{\star}[t,\tau], p_{i,c}^{{\rm u},\star}[t, \tau], \phi_{m}^{(\ell), \star}[t, \tau]\}$.
\end{algorithmic}
\end{algorithm}

\vspace{-0.9 em}
\section{Proposed Solution} \label{Solution}
Problem (\ref{Problem_1}) is a mixed-integer non-convex optimization problem with two time scales of decision making: (i) eMBB variables \(\{\phi_{m}^{(\ell)}[t], \alpha_{i,c}[t], p_{i,c}^{\rm{e}}[t]\}\), determined at the beginning of each time slot \(t\); and (ii) URLLC (target) variables \(\{\phi_{m}^{(\ell)}[t, \tau], \beta_{i,c}[t,\tau], \rho_{i,c}[t,\tau], p_{i,c}^{\rm{u}}[t,\tau]\}\), updated at each mini-slot \((t, \tau)\) to accommodate the instantaneous URLLC arrivals by puncturing RBs initially allocated to eMBB users. The SIM phase responses \(\{\phi_{m}^{(\ell)}\}\) are configurable at both time scales and influence both services through SIM transfer matrix \(\mathbf{\Theta}\). 
Motivated by this two-timescale structure and the separability of the objective function in~\eqref{Problem_1}, we decompose (\ref{Problem_1}) into two coupled subproblems: an \emph{EE subproblem for eMBB} solved once per time slot, and an \emph{EE subproblem for URLLC} solved at each mini-slot. Specifically, at time slot \(t\), for given feasible URLLC variables, we solve
\begin{equation} \label{Prob_eMBB} 
\begin{aligned} 
     &\max_{\substack{
        \{\phi_{m}^{(\ell)}[t], \alpha_{i, c}[t], p_{i, c}^{\rm{e}}[t]\} 
    }} 
    EE^{\rm{e}}[t] \\
    \text{s.t.} \:& \text{C1}, \text{C6}, \text{C10}, \text{C11}, \text{C13}, \text{C16}.
\end{aligned}
\end{equation}

For each mini-slot \((t,\tau)\), given feasible eMBB variables, we solve
% \footnote{Since no inter-temporal coupling constraints are imposed across mini-slots in problem \eqref{Problem_1}, the EE subproblem for URLLC users can be solved independently at each mini-slot $(t, \tau)$.}
\begin{equation}\label{Prob_URLLC} 
\begin{aligned} 
     &\max_{\substack{
        \{\phi_{m}^{(\ell)}[t, \tau], \beta_{i, c}[t, \tau]\} \\
        \{\rho_{i, c}[t, \tau], p_{i, c}^{\rm{u}}[t, \tau]\}
    }} 
    EE^{\rm{u}}[t, \tau] \\
    \text{s.t.} \:& \text{C1--C5}, \text{C7--C10}, \text{C12}, \text{C14--C15}, \text{C17}.
\end{aligned}
\end{equation}
where \text{C1} is enforced on the minimum data rate of eMBB users because puncturing and mini-slot SIM adaptation can reduce the achievable data rate of eMBB users within the time slot. The two subproblems are therefore coupled through \text{C1} and \text{C10}, as well as the SIM transfer matrix \(\mathbf{\Theta}\). 

To jointly solve the coupled subproblems, we adopt a two-timescale AO framework. Specifically, at each time slot \(t\), the eMBB EE subproblem is first solved. Then, for each mini-slot \((t, \tau)\), the URLLC EE subproblem is solved using the updated eMBB variables. The resulting URLLC decisions are then used as inputs for the subsequent iteration. Since each subproblem is optimized locally with respect to its own variables while keeping the others fixed, the objective value of problem \eqref{Problem_1} is non-decreasing across iterations. Moreover, since the objective is upper bounded by transmit power and resource constraints, the proposed two-timescale algorithm, given in Algorithm~\ref{alg:overall}, is guaranteed to converge to a locally optimal solution of problem \eqref{Problem_1}. This iterative procedure continues across time slots. In the following, we describe the solution approach for subproblems \eqref{Prob_eMBB} and \eqref{Prob_URLLC}.

\vspace{-1 em}
\subsection{Solving the eMBB EE Subproblem \eqref{Prob_eMBB}} \label{Solve_eMBB}
Obtaining the optimal solution to problem \eqref{Prob_eMBB} is challenging due to: i) the fractional objective function and ii) the coupling among RB assignment, power allocation, and SIM phase shifts in both the objective function and constraints. To address these difficulties, we first convert problem \eqref{Prob_eMBB} into a more tractable form via the following two steps.
\newline
\textit{Step (a):} To handle the fractional objective function in \eqref{Prob_eMBB} at time slot $t$, we apply Dinkelbach's method \cite{Dinkelbach}. Specifically, we introduce an auxiliary parameter $\eta^{\rm{e}}[t] \ge 0$ to transform the objective function of \eqref{Prob_eMBB} into an equivalent subtractive form as $\sum_{i\in\mathcal{U}^{\rm{e}}} r_{i}^{\rm{e}}[t] - \eta^{\rm{e}}P_{\mathrm{tot}}^{\rm{e}}[t]$. According to Dinkelbach's method \cite{Dinkelbach}, for $\sum_{i\in\mathcal{U}^{\rm{e}}} r_{i}^{\rm{e}}[t] \geq 0$ and $P_{\mathrm{tot}}^{\rm{e}}[t] > 0$, the optimal solution $\{\alpha_{i,c}^{\star}[t],p_{i,c}^{{\rm e},\star}[t], {\phi}_{m}^{(\ell), \star}[t]\}$ to problem \eqref{Prob_eMBB} is achieved if and only if there exists $\eta^{\rm{e}, \star}[t]$ such that $\max \big(\sum_{i\in\mathcal{U}^{\rm{e}}} r_{i}^{\rm{e}}[t] - \eta^{\rm{e},\star}[t] P_{\mathrm{tot}}^{\rm{e}}[t]\big) = 0$, where $\eta^{\rm{e}, \star}[t]$ denotes the optimal EE in problem \eqref{Prob_eMBB}. Therefore, for a given \(\eta^{{\rm{e}}, (j)}[t]\) at iteration $j$ of the Dinkelbach's method, problem \eqref{Prob_eMBB} is rewritten as
\begin{equation} \label{Prob_eMBB_2} 
\begin{aligned} 
     &\max_{\substack{
    \{\phi_{m}^{(\ell)}[t], \alpha_{i,c}[t], p_{i,c}^{\rm{e}}[t]\}
    }} \:
    \widetilde{EE}^{\rm{e}}[t] = \sum_{i\in\mathcal{U}^{\rm{e}}} r_{i}^{{\rm{e}}}[t] - \eta^{{\rm{e}}, (j)}[t]\, P_{\mathrm{tot}}^{{\rm{e}}}[t] \\
    &\text{s.t.} \ \text{C1}, \text{C6}, \text{C10}, \text{C11}, \text{C12}, \text{C15}.
\end{aligned}
\end{equation}
After solving \eqref{Prob_eMBB_2} for the given \(\eta^{{\rm{e}}, (j)}[t]\), the Dinkelbach factor is updated as \cite{Dinkelbach}
\begin{equation} \label{eMBB_eta}
\eta^{{\rm{e}}, (j+1)}[t] = 
\frac{\sum_{i\in\mathcal{U}^{\rm{e}}} r_{i}^{{\rm{e}}}[t]}
{P_{\mathrm{tot}}^{{\rm{e}}}[t]}.
\end{equation}
The iterations continue until convergence.
% which corresponds to the achieved EE at iteration $j$
%, i.e., $\big(\sum_{i\in\mathcal{U}^{{\rm{e}}}} r_{i}^{{\rm{e}}}[t] - \eta^{{\rm{e}},(j)}[t] P_{\mathrm{tot}}^{{\rm{e}}}[t] \rightarrow  0 \big)$
\newline
\textit{Step (b):} To address the coupling among RB allocation \(\{\alpha_{i,c}[t]\}\), power allocation \(\{p_{i,c}^{\rm{e}}[t]\}\), and SIM phase shifts \(\{\phi_{m}^{(\ell)}[t]\}\), we decompose problem \eqref{Prob_eMBB_2} into three subproblems, namely, RB allocation, power allocation, and SIM phase shifts. We then adopt an AO procedure. Specifically, starting from an initial feasible solution, the variables are updated iteratively in three stages: (i) RB allocation is optimized for given power allocation and SIM phase shifts; (ii) power allocation is updated for given RB allocation and SIM phase shifts; and (iii) SIM phase shifts are refined for given RB and power allocation. The updated variables at each stage are used in the subsequent stages, and this process is repeated until convergence, i.e., when the variation of the objective value between successive iterations falls below a predefined threshold or a maximum number of iterations is reached.

\subsubsection{RB allocation subproblem} The RB allocation subproblem is expressed as
\begin{equation} \label{Prob_eMBB_2_1} 
\begin{aligned} 
     &\max_{\substack{\{\alpha_{i,c}[t]\}}} \:
   \widetilde{EE}^{\rm{e}}[t]\quad \text{s.t.}\ \text{C1}, \text{C6}, \text{C10}, \text{C13}.
\end{aligned}
\end{equation}
This is an integer nonlinear programming problem and can be solved using MATLAB/CVX with MOSEK solver \cite{MOSEK}. 

\subsubsection{Power allocation subproblem} The power allocation subproblem is formulated as
\begin{equation} \label{Prob_eMBB_2_2} 
\begin{aligned} 
     &\max_{\substack{\{p_{i,c}^{\rm{e}}[t]\}}} \:
    \widetilde{EE}^{\rm{e}}[t]\quad \text{s.t.} \ \text{C1}, \text{C10}, \text{C16}.
\end{aligned}
\end{equation}
This problem is convex and can be solved using CVX \cite{Boyd2004}.

\subsubsection{SIM phase shift subproblem}
Since RB assignment and power allocation are fixed, maximizing \(\widetilde{EE}^{\rm{e}}[t]\) is equivalent to maximizing the achievable sum data rate. Therefore, we solve
\begin{equation}\label{Prob_eMBB_2_3}
\begin{aligned}
&\max_{\{\phi_{m}^{(\ell)}[t]\}}  \sum_{i\in\mathcal{U}^{\rm{e}}} r_{i}^{\rm{e}}[t] \quad  {\rm s.t.}\ \text{C1}, \text{C11}.
\end{aligned}
\end{equation}
To facilitate tractable optimization, constraint C1 is incorporated into the objective function via a penalty term, yielding
\begin{equation}\label{Prob_eMBB_2_3_penalty}
\begin{aligned}
&\max_{\{\phi_{m}^{(\ell)}[t]\}}  
\sum_{i\in\mathcal{U}^{\rm{e}}} r_{i}^{\rm{e}}[t] 
- \zeta_1 \sum_{i\in\mathcal{U}^{\rm{e}}} \left[r_{i}^{\rm{min}} - r_{i}^{\rm{e}}[t]\right]^+ 
\quad \text{s.t.}\ \text{C11},
\end{aligned}
\end{equation}
where $\zeta_1 \ge 0$ is a penalty parameter.
We then employ projected gradient ascent (PGA) \cite{An_2025} with Wirtinger gradients. Starting from a feasible initial point satisfying $|\phi_{m}^{(\ell)}[t]|=1$, each iteration consists of a gradient ascent step followed by a projection onto the unit-modulus constraint set, ensuring $|\phi_{m}^{(\ell)}[t]|=1$ for all $m$ and $\ell$.

By iteratively solving the subproblems \eqref{Prob_eMBB_2_1}, \eqref{Prob_eMBB_2_2}, and \eqref{Prob_eMBB_2_3_penalty}, a suboptimal solution to problem \eqref{Prob_eMBB_2} at time slot \(t\) is obtained. The corresponding procedure is summarized in Algorithm \ref{alg:eMBB}.

\begin{algorithm}[t!]
\SetAlCapFnt{\footnotesize}
\SetAlCapNameFnt{\footnotesize}
\caption{EE optimization for eMBB at time slot $t$} \label{alg:eMBB}
\footnotesize
\begin{algorithmic}[1]
\STATE \textbf{Input:} Maximum number of iterations $J_{\max}$ and $N_{\max}$, tolerances \\ $\epsilon_1$ and $\epsilon_2$, and initial SIM phase $\phi_{m}^{(\ell)}[t,0]$.
\STATE Set $\phi_{m}^{(\ell), (0)}[t] \gets \phi_{m}^{(\ell)}[t,0]$, $j\gets 0$, and $\eta^{{\rm e},(0)}[t]\gets 0$.
\REPEAT
    \STATE Initialize $\{\alpha_{i,c}^{(0)}[t], p_{i,c}^{{\rm e},(0)}[t]\}$ and set $n\gets 0$.
    \REPEAT
        \STATE Solve subproblem \eqref{Prob_eMBB_2_1} to obtain $\alpha_{i,c}^{(n+1)}[t]$.
        \STATE Solve subproblem \eqref{Prob_eMBB_2_2} to obtain $p_{i,c}^{{\rm e},(n+1)}[t]$.
        \STATE Solve subproblem \eqref{Prob_eMBB_2_3_penalty} to obtain $\phi_{m}^{(\ell), (n+1)}[t]$.
        \STATE Set $n\gets n+1$.
    \UNTIL{$|\widetilde{EE}^{{\rm e},(n)}[t]-\widetilde{EE}^{{\rm e},(n-1)}[t]|<\epsilon_1$ or $n=N_{\max}$}
    \STATE Update $\eta^{{\rm e},(j+1)}[t]$ via \eqref{eMBB_eta}, and set $j\gets j+1$.
\UNTIL{$|\eta^{{\rm e},(j)}[t]-\eta^{{\rm e},(j-1)}[t]|\le\epsilon_2$ or $j=J_{\max}$}
\STATE Set $\widetilde{EE}^{\rm e}[t]=\sum_{i\in\mathcal{U}^{\rm e}} r_i^{\rm e}[t]-\eta^{{\rm e},(j)}[t]P_{\rm tot}^{\rm e}[t]$.
\STATE Obtain $\!\alpha_{i,c}^{\star}[t]\gets \alpha_{i,c}^{(n)}[t]$, $p_{i,c}^{{\rm e},\star}[t]\gets p_{i,c}^{{\rm e},(n)}[t], \phi_{m}^{(\ell), \star}[t]\gets \phi_{m}^{(\ell), (n)}[t]$.
\STATE \textbf{Output:} $\{\alpha_{i,c}^{\star}[t],p_{i,c}^{{\rm e},\star}[t], \phi_{m}^{(\ell), \star}[t]\}$, and $\widetilde{EE}^{\rm e}[t]$.
\end{algorithmic}
\end{algorithm}

\vspace{-1 em}
\subsection{Solving the URLLC EE Subproblem \eqref{Prob_URLLC}}
Similar to problem \eqref{Prob_eMBB}, problem~\eqref{Prob_URLLC} is a mixed-integer non-convex optimization problem due to: i) the fractional objective function and ii) the coupling among RB assignment, power allocation, and SIM phase shifts in both the objective function and constraints. In addition, the URLLC EE problem \eqref{Prob_URLLC} introduces three further challenges: iii) the probabilistic reliability constraint \text{C2}, iv) the total transmit power consumption of URLLC users in \eqref{total_P_URLLC}, and v) the long-term average AoI constraint \text{C4}. The difficulties in (i) and (ii) are addressed using the same steps as in Subsection~\ref{Solve_eMBB}, i.e., we first apply Dinkelbach's transformation to convert the fractional objective into a subtractive form and then adopt an AO procedure over RB assignment, power allocation, and SIM phase shifts. To deal with the probabilistic nature of constraint C2 in \eqref{Prob_URLLC}, following the approach in \cite{Bairagi2021}, we transform constraint C2 into a deterministic constraint as
\begin{equation*}
    \widehat{\text{C2}}: r_{i}^{\rm{u}}[t, \tau] - L_{i} F_{{N}_{i}[t, \tau]}^{-1}(\gamma_{i}^{\rm{Rel}}) \geq 0, \: \forall i \in \mathcal{U}^{\rm{u}}, \forall \tau \in \mathcal{I},  \forall t \in \mathcal{T},
\end{equation*}
where $F_{{N}_{i}[t, \tau]}^{-1}(\cdot)$ denotes the inverse cumulative distribution function of the random variable ${N}_{i}[t, \tau]$. Additionally, to tackle the total transmit power consumption of URLLC users in~\eqref{total_P_URLLC}, we introduce an auxiliary binary variable \(z_{i,c}[t,\tau]\in\{0,1\}\) to represent the activation of RB \(c\) for URLLC user \(i\) in mini-slot \((t,\tau)\), i.e., \(z_{i,c}[t,\tau]=\beta_{i,c}[t,\tau]\rho_{i,c}[t,\tau]\), and impose the following linear constraints:
\begin{equation}\label{z_or_lin}
\begin{aligned}
    \text{C18.1: } z_{i,c}[t,\tau]& \leq \beta_{i,c}[t,\tau],\\
    \text{C18.2: } z_{i,c}[t,\tau]& \leq \rho_{i,c}[t,\tau],\\
    \text{C18.3: } z_{i,c}[t,\tau]& \ge \beta_{i,c}[t,\tau]+\rho_{i,c}[t,\tau] - 1.
\end{aligned}
\end{equation}

By taking these transformations, problem \eqref{Prob_URLLC} is decomposed into three subproblems: i) RB allocation, (ii) power allocation, and (iii) SIM phase shift optimization, which are solved alternately as summarized in Algorithm~\ref{alg:URLLC}.
\subsubsection{RB allocation subproblem}\label{DPP} 
The RB allocation subproblem is formulated as     
\begin{equation} \label{Prob_URLLC_1_1} 
\begin{aligned} 
     &\max_{\substack{\{z_{i,c}[t,\tau]\},\\\{\beta_{i, c}[t, \tau], \rho_{i, c}[t, \tau]\}}} \!\!\!\!\!\!\!
    \widetilde{EE}^{\rm{u}}[t, \tau] = \!\!\!\sum_{i\in\mathcal{U}^{\rm{u}}} \!\! r_{i}^{{\rm{u}}}[t, \tau] \!\!-\!\! \eta^{{\rm{u}}, (j)}[t, \tau] P_{\mathrm{tot}}^{{\rm{u}}}[t, \tau]\\
    & \text{s.t.} \: \text{C1}, \widehat{\text{C2}}, \text{C3--C5}, \text{C7--C10}, \text{C14--C15}, {\text{C18.1--C18.3}},
\end{aligned}
\end{equation}
where \(\eta^{{\rm{u}}, (j)}[t,\tau]\ge 0\) is the Dinkelbach factor in iteration $j$ of Dinkelbach method, updated iteratively according to
\begin{equation} \label{URLLC_eta}
\eta^{{\rm{u}}, (j+1)}[t, \tau] = 
\frac{\sum_{i\in\mathcal{U}^{\rm{u}}} r_{i}^{{\rm{u}}}[t, \tau]}
{P_{\mathrm{tot}}^{{\rm{u}}}[t, \tau]},
\end{equation}
until convergence.
Problem \eqref{Prob_URLLC_1_1} is particularly challenging due to the long-term average AoI constraint \text{C4}, which cannot be enforced within a single mini-slot. To handle C4, we adopt the Lyapunov drift-plus-penalty (DPP) framework \cite{neely2010stochastic}, which converts this constraint into a virtual queue stability condition. Specifically, a virtual queue $U_{i}[t, \tau]$ is associated with each target $i \in \mathcal{U}^{\rm{u}}$ and updated as
\begin{equation}\label{eq:virtual_queue_update}
    U_{i}[t, \tau+1] \;=\; [\, U_{i}[t, \tau] \;+\; \delta_{i}[t, \tau+1] - \Delta_{i}^{\max}\,]^+,
\end{equation}
with $U_{i}[1, 0] = 0$ and $[\cdot]^+ = \max\{\cdot,0\}$. Here, $\delta_{i}[t, \tau+1]$ acts as a (virtual) arrival rate and $\Delta_{i}^{\max}$ is interpreted as a (virtual) service rate. Under the Lyapunov optimization, stabilizing virtual queues $\{U_{i}[t, \tau]\}$ directly ensures that C4 is met because the stability of queues implies that the time-average arrival rate does not exceed the service rate, i.e.,
\begin{equation*}\label{eq:aoi_avg_constraint}
\lim_{T\to\infty}\lim_{I\to\infty}\frac{1}{T\times I}\sum_{t=1}^{T}\sum_{\tau=0}^{I-1}\mathbb{E}\!\left[\delta_{i}[t, \tau+1]\right]\le \Delta_{i}^{\max},\: \forall i\in\mathcal{U}^{\rm{u}}.
\end{equation*}
To characterize stability and derive an online control rule, we define the quadratic Lyapunov function $L(\mathbf U[t, \tau])$ as
\begin{equation}\label{eq:Lyapunov}
    L(\mathbf U[t, \tau]) \;=\; \frac{1}{2}\sum_{i\in \mathcal{U}^{\rm{u}}} U_{i}[t, \tau]^2
\end{equation}
and the conditional Lyapunov drift $\Delta L[t, \tau]$ as
\begin{equation}\label{eq:drift}
    \Delta L[t, \tau] \;=\; \mathbb{E}\!\left[L(\mathbf U[t, \tau+1]) - L(\mathbf U[t, \tau]) \,\middle|\, \mathcal{H}[t, \tau]\right],
\end{equation}
where $\mathcal{H}[t, \tau]=\{U_{i}[t, \tau], \delta_{i}[t, \tau]\}_{i\in \mathcal{U}^{\rm{u}}}$ denotes the current network state at the beginning of mini-slot $(t, \tau)$. According to the DPP principle \cite{neely2010stochastic}, minimizing $\Delta L[t, \tau]$ over time drives the virtual queues toward stability. A control policy that jointly enforces queue stability and optimizes the objective in \eqref{Prob_URLLC_1_1} is obtained by solving
\begin{equation} \label{Prob_URLLC_1_1_1} 
\begin{aligned} 
     &\max_{\substack{
    \{\beta_{i, c}[t, \tau], \rho_{i, c}[t, \tau]\}
    }} \:\:
    V \: \widetilde{EE}^{\rm{u}}[t, \tau] - \Delta L[t, \tau]\\
    & \text{s.t.} \: \text{C1}, \widehat{\text{C2}}, \text{C3}, \text{C5}, \text{C7--C10}, \text{C13--C14}, {\text{C18.1--C18.3}},
\end{aligned}
\end{equation}
where $V\geq 0$ controls the trade-off between queue backlog (AoI satisfaction) and the objective function value. To obtain a tractable problem, we derive an upper bound on $\Delta L[t, \tau]$. Using the inequality $(x^+)^2 \le x^2$, for all $i\in \mathcal{U}^{\rm{u}}$, we have 
\begin{equation} \label{Approx_1}
\begin{aligned} 
U_{i}^{2}[t, \tau+1] &= {\big[\, U_{i}[t, \tau] \;+\; \delta_{i}[t, \tau+1] - \Delta_{i}^{\max}\,\big]^+}^{2}\\
& \leq (U_{i}[t, \tau]  + \delta_{i}[t, \tau+1] - \Delta_{i}^{\max} )^{2} \\
& = U_{i}^{2}[t, \tau] + (\delta_{i}[t, \tau+1] - \Delta_{i}^{\max})^{2} + \\
&  \quad 2 U_{i}[t, \tau] (\delta_{i}[t, \tau+1] - \Delta_{i}^{\max}).
\end{aligned}
\end{equation}
Substituting (\ref{Approx_1}) into \eqref{eq:drift} yields an upper bound as
\begin{equation}
\begin{aligned} \label{Approx_2}
\Delta L[t, \tau] &\leq \frac{1}{2} \mathbb{E}\big[\sum_{i\in\mathcal{U}^{\rm{u}}}  \left(U_{i}^{2}[t, \tau+1] - U_{i}^{2}[t, \tau]\right) |\, \mathcal{H}[t, \tau] \big]\\
&\leq Z + \sum_{i\in\mathcal{U}^{\rm{u}}}  U_{i}[t, \tau] \left(\delta_{i}[t, \tau+1] - \Delta_{i}^{\max}\right),
\end{aligned}
\end{equation}
where $Z$ is a finite constant satisfying
$Z  \ge  \frac{1}{2}\sum_{i\in\mathcal{U}^{\rm{u}}} (\delta_{i}[t, \tau+1] - \Delta_{i}^{\max})^2$.
Since $Z$ is independent of the current control variables, it can be dropped when forming the per-mini-slot maximization. According to the definition of $\delta_{i}[t, \tau+1]$ in \eqref{AOI_def}, we can express $\delta_{i}[t, \tau+1]$ as a function of detecting decisions as
\begin{equation} \label{aoi_rec}
    \delta_{i}[t, \tau+1] \;=\; \delta_{i}[t, \tau] + 1 - \rho_{i}[t, \tau] \delta_{i}[t, \tau],
\end{equation}
where $\rho_{i}[t, \tau] = \sum_{c \in \mathcal{C}} \rho_{i, c}[t, \tau] \in \{0,1\}$ indicates whether target $i$ is detected in mini-slot $(t, \tau)$. Substituting \eqref{aoi_rec} into \eqref{Approx_2} and discarding constant $Z$, the drift term reduces to
\begin{equation*}
\Delta L[t, \tau] \;\le\;\sum_{i\in\mathcal{U}^{\rm{u}}} U_{i}[t, \tau]\big(1 + \delta_{i}[t, \tau] - \rho_{i}[t, \tau] \,\delta_{i}[t, \tau] - \Delta_{i}^{\max}\big).
\end{equation*}
Therefore, problem \eqref{Prob_URLLC_1_1_1} can be rewritten as
\begin{equation} \label{Prob_URLLC_1_1_2} 
\begin{aligned} 
     &\max_{\substack{\{z_{i, c}[t, \tau]\},\\
    \{\beta_{i, c}[t, \tau], \rho_{i, c}[t, \tau]\} }} 
    V\: \widetilde{EE}^{\rm{u}}[t, \tau]\\
    &\qquad - \sum_{i\in\mathcal{U}^{\rm{u}}} U_{i}[t, \tau]\big(1 + \delta_{i}[t, \tau] - \rho_{i}[t, \tau] \delta_{i}[t, \tau] - \Delta_{i}^{\max}\big)\\
    & \text{s.t.} \: \text{C1}, \widehat{\text{C2}}, \text{C3}, \text{C5}, \text{C7--C10}, \text{C14--C15}, {\text{C18.1--C18.3}}.
\end{aligned}
\end{equation}
This is an integer nonlinear programming problem and can be solved using MATLAB/CVX with MOSEK solver \cite{MOSEK}.

\subsubsection{Power allocation subproblem} The power allocation subproblem is expressed as
\begin{equation} \label{Prob_URLLC_1_2} 
\begin{aligned} 
     &\max_{\substack{\{p_{i,c}^{\rm{u}}[t, \tau]\}}} \:
    \widetilde{EE}^{\rm{u}}[t, \tau]  \quad \text{s.t.} \quad \widehat{\text{C2}}, \text{C3}, \text{C5}, \text{C10}, \text{C17}.
\end{aligned}
\end{equation}
Problem \eqref{Prob_URLLC_1_2} is convex and can be efficiently solved using CVX~\cite{Boyd2004}.

\subsubsection{SIM phase shift subproblem} Since RB assignment and power allocation are fixed, maximizing $\widetilde{EE}^{\rm{u}}[t, \tau]$ is equivalent to maximizing the achievable sum data rate. Therefore, we solve 
\begin{equation}\label{Prob_URLLC_1_3}
\begin{aligned}
    &\max_{\{\phi_{m}^{(\ell)}[t, \tau]\}}
    \sum_{i\in\mathcal{U}^{\rm{u}}} r_{i}^{\rm{u}}[t, \tau] \quad {\rm s.t.}\ \text{C1}, \widehat{\text{C2}}, \text{C3}, \text{C5}, \text{C12}.
\end{aligned}
\end{equation}
Similar to problem \eqref{Prob_eMBB_2_3_penalty}, to facilitate tractable optimization, constraints $\text{C1}, \widehat{\text{C2}}, \text{C3}, \text{C5}$ are incorporated into the objective function via penalty terms. Then, we employ PGA with Wirtinger gradients \cite{An_2025} for updating the SIM phase shifts.

\begin{algorithm}[t!]
\SetAlCapFnt{\footnotesize}
\SetAlCapNameFnt{\footnotesize}
\caption{EE optimization for URLLC at mini-slot $(t,\tau)$}
\label{alg:URLLC}

\footnotesize
\begin{algorithmic}[1]
\STATE \textbf{Input:} Maximum number of iterations $J_{\max}$ and $N_{\max}$, tolerances \\ $\epsilon_1$ and $\epsilon_2$, and SIM phase $\phi_{m}^{(\ell), \star}[t]$ at time slot $t$.
\STATE Set $\phi_{m}^{(\ell), (0)}[t, \tau] \gets \phi_{m}^{(\ell), \star}[t]$, $j\gets 0$, and $\eta^{{\rm u},(0)}[t,\tau]\gets 0$.
\REPEAT
    \STATE Initialize $\{\beta_{i,c}^{(0)}[t,\tau],\rho_{i,c}^{(0)}[t,\tau],p_{i,c}^{{\rm u},(0)}[t,\tau]\}$ and set $n\gets 0$.
    \REPEAT
        \STATE Solve subproblem \eqref{Prob_URLLC_1_1_2} to obtain $\beta_{i,c}^{(n+1)}[t,\tau]$ and $\rho_{i,c}^{(n+1)}[t,\tau]$.
        \STATE Solve subproblem \eqref{Prob_URLLC_1_2} to obtain $p_{i,c}^{{\rm u},(n+1)}[t,\tau]$.
        \STATE Solve subproblem \eqref{Prob_URLLC_1_3} to obtain $\phi_{m}^{(\ell), (n+1)}[t, \tau]$.
        \STATE Set $n\gets n+1$.
    \UNTIL{$|\widetilde{EE}^{{\rm u},(n)}[t,\tau]-\widetilde{EE}^{{\rm u},(n-1)}[t,\tau]|<\epsilon_1$ or $n=N_{\max}$}
    \STATE Update $\eta^{{\rm u},(j+1)}[t,\tau]$ via \eqref{URLLC_eta}, and set $j\gets j+1$.
\UNTIL{$|\eta^{{\rm u},(j)}[t,\tau]-\eta^{{\rm u},(j-1)}[t,\tau]|\le\epsilon_2$ or $j=J_{\max}$}
\STATE Set $\widetilde{EE}^{\rm u}[t,\tau]
=\sum_{i\in\mathcal{U}^{\rm u}} r_i^{\rm u}[t,\tau]
-\eta^{{\rm u},(j)}[t,\tau]P_{\rm tot}^{\rm u}[t,\tau]$.
\STATE Obtain $\beta_{i,c}^{\star}[t,\tau]\gets \beta_{i,c}^{(n)}[t,\tau]$, $\rho_{i,c}^{\star}[t,\tau]\gets \rho_{i,c}^{(n)}[t,\tau]$, $p_{i,c}^{{\rm u},\star}[t,\tau]\gets p_{i,c}^{{\rm u},(n)}[t,\tau]$, and $\phi_{m}^{(\ell),\star}[t,\tau]\gets \phi_{m}^{(\ell),(n)}[t, \tau]$.
\STATE \textbf{Output:} $\{\beta_{i,c}^{\star}[t,\tau],\rho_{i,c}^{\star}[t,\tau],p_{i,c}^{{\rm u},\star}[t,\tau]\}$, $\phi_{m}^{(\ell), \star}[t,\tau]$, and $\widetilde{EE}^{\rm u}[t,\tau]$.
\end{algorithmic}
\end{algorithm}

\vspace{-1 em}
\subsection{Computational Complexity of Algorithm \ref{alg:overall}}
Algorithm~\ref{alg:overall} solves problem \eqref{Problem_1} by decomposing it into an eMBB EE optimization at each time slot and a URLLC EE optimization at each mini-slot. For the eMBB EE subproblem \eqref{Prob_eMBB}, the dominant complexity stems from the updates of RB allocation, transmit power, and SIM phase shifts in Algorithm~\ref{alg:eMBB}. The RB allocation \eqref{Prob_eMBB_2_1} and power allocation \eqref{Prob_eMBB_2_2} subproblems are solved using CVX via an interior-point method (IPM)~\cite{Boyd2004}. Since the complexity of IPM scales cubically with the number of decision variables, and both subproblems involve \(U^{\rm{e}}C\) variables, their complexity is $\mathcal{O}\!\left(I_{\rm IPM}(U^{\rm{e}}C)^3\right)$, where \(I_{\rm IPM}=\frac{\log(N_{\rm{e}}/(i^0\Lambda))}{\log(\varsigma)}\) denotes the number of IPM iterations, with \(N_{\rm{e}}=2U^{\rm{e}}C+2U^{\rm{e}}+C+2\) being the total number of constraints in \eqref{Prob_eMBB_2_1} and \eqref{Prob_eMBB_2_2}, \(i^0\) the initial accuracy parameter, \(0<\Lambda\ll1\) the stopping tolerance, and \(\varsigma>1\) the accuracy update factor~\cite{Boyd2004}. The SIM phase-shift update via PGA over \(ML\) variables involves Wirtinger gradient computation and projection onto the unit circle at each iteration, with complexity \(\mathcal{O}(N_{\rm PGA}\,U^{\rm{e}}C\,LM^{2})\), where $N_{\rm PGA}$ is the number of PGA iterations.
Therefore, the overall per-time-slot complexity of Algorithm~\ref{alg:eMBB} is $\mathcal{O}\left(J_{\rm{max}} N_{\rm{max}}[I_{\rm IPM}(U^{\rm{e}}C)^3 + N_{\rm PGA}\,U^{\rm{e}}C\,LM^{2}]\right)$, where $J_{\rm{max}}$ and $N_{\rm{max}}$ denote the numbers of Dinkelbach and AO iterations, respectively.
Similarly, the per-mini-slot complexity of Algorithm~\ref{alg:URLLC} for solving the URLLC EE subproblem \eqref{Prob_URLLC} is $\mathcal{O}(J_{\rm{max}} N_{\rm{max}}[I_{\rm IPM}(U^{\rm{u}}C)^{3}+N_{\rm PGA}\,U C\,LM^{2}])$, where $U = U^{\rm{e}} + U^{\rm{u}}$ denotes the total number of users.

\vspace{-1 em}
\section{Numerical Results} \label{results}
In this section, we evaluate the performance of the proposed algorithm. We consider a downlink SIM-enabled ISAC system. The BS is deployed at height $H_{\rm{BS}}=10$m, and is equipped with a SIM consisting of $L=3$ layers, each with $M=36$ meta-atoms \cite{Zhang_2025}. The inter-element spacing is $\lambda_{c}/2$, and the SIM thickness is $5\lambda_{c}$, with equally spaced layers \cite{Zhang_2025}. User locations are generated in polar coordinates with distance uniformly distributed in $[5, 50]$m.

Unless otherwise stated, the default parameters are as follows. We assume that the BS uses $C=25$ RBs, each with a bandwidth of $ B=180$ kHz, at a carrier frequency of $f_{c}=5$GHz \cite{Zhang_2025}. We consider four users for each service, i.e., $U^{\rm{e}}=U^{\rm{u}}=4$. For URLLC users \cite{TS38211, Prathyusha2022, Alsenwi2021}, the transmission duration is $T_{d} = 5 \times 10^{-4}$s, with a short packet size of $32$ bytes. The blocklength is calculated by $(B \times T_{d}) / (2I)$. Also, the desired decoding error probability is $\epsilon = 10^{-5}$, and the reliability requirement for each URLLC user $i$ is set to $\gamma_{i}^{\rm{Rel}}= 0.99999$. The average arrival rate of URLLC users is $\lambda_{i}^{\rm{u}}=0.5$. 

URLLC users also act as sensing targets with AoI thresholds $\Delta_{i}^{\rm{max}}=[1,2,3, 4]$, and the minimum sensing beampattern gain is set to $\Gamma^{\rm{th}}=-25$dBm. The background noise and path-loss exponent are set to -95 dBm and $\alpha=3.5$, respectively. The maximum transmit power is $P^{\rm{max}}=5$W. In all scenarios, the numerical results are averaged over 50 independent snapshots; for each snapshot, the system is simulated over $T=6$ time slots, and each time slot is divided into $I=7$ mini-slots spanning two OFDM symbols \cite{TS38211, Prathyusha2022, Alsenwi2021}. 

\begin{figure}[t!]	
    \centering
    \includegraphics[width=8 cm, height=4 cm]{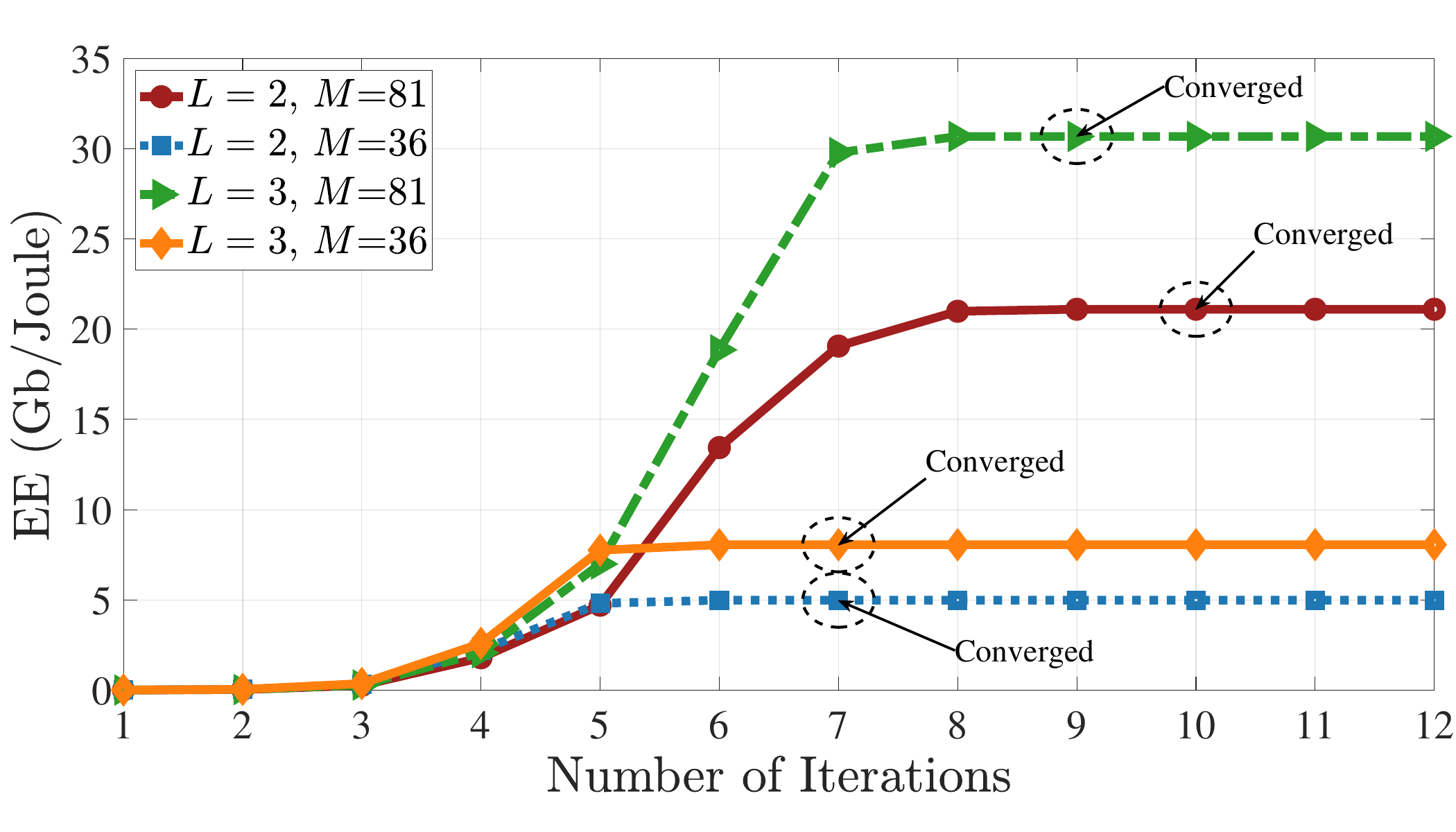}
    \vspace{-5 pt}
    \caption{Convergence of the proposed algorithm. \label{fig:Convergence}} 
\end{figure}

\begin{figure}[t!]	
    \centering
    \includegraphics[width=8 cm, height=4 cm]{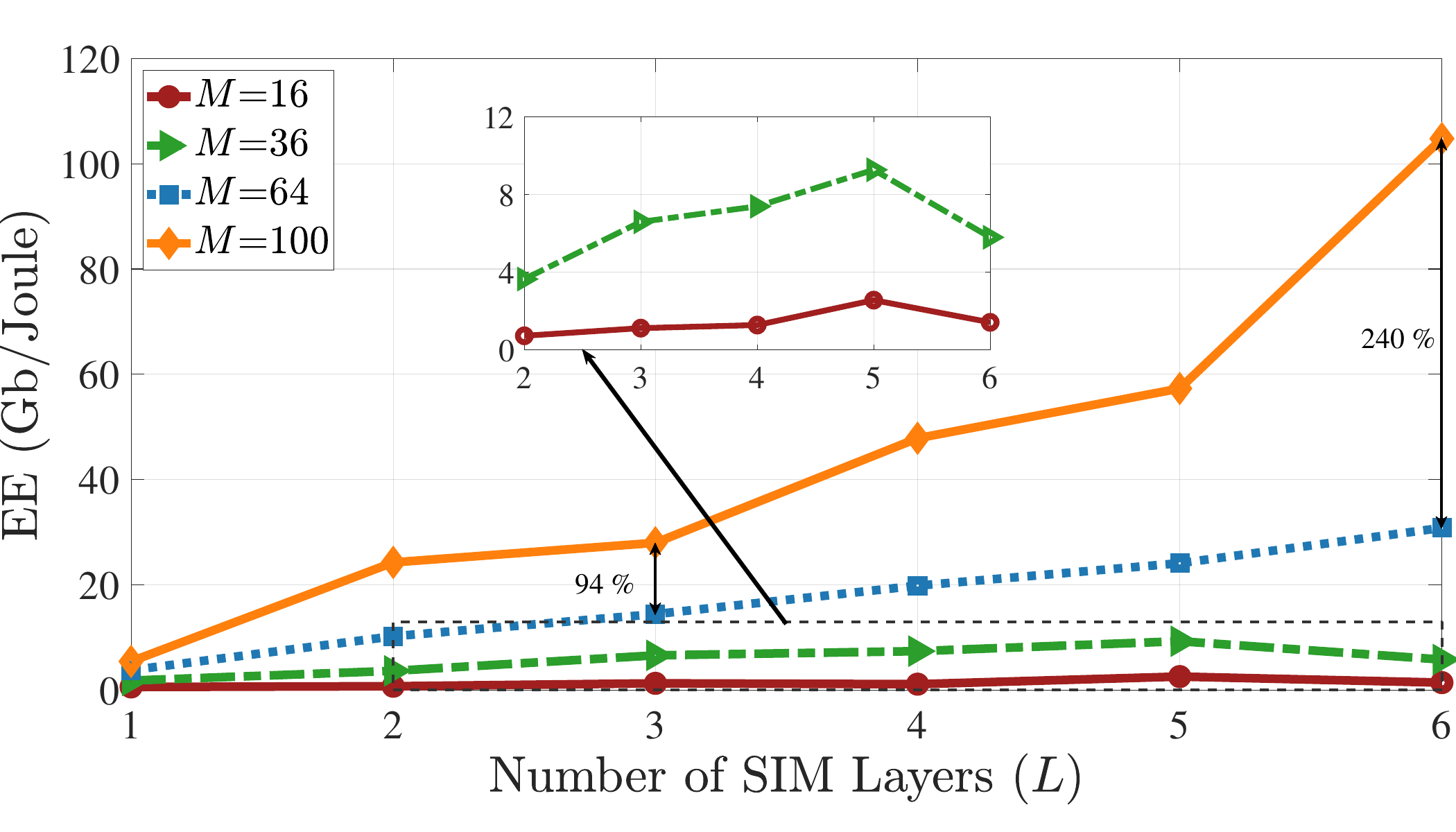}
    \vspace{-5 pt}
    \caption{EE versus the number of SIM layers.\label{fig:Layers}}
\end{figure}

\vspace{-0.8 em}
\subsection{Convergence of Our Proposed Algorithm}
Fig.~\ref{fig:Convergence} illustrates the convergence behavior of the proposed iterative algorithm versus the number of iterations for different numbers of SIM layers and meta-atoms per layer. To generate this figure, the minimum data rate requirement for eMBB users is set to $r_{i}^{\rm{min}}=1$Mbps, and the maximum tolerable latency for URLLC users is set to $T_{i}^{\rm{max}} = 1.5$ms. It can be observed that the EE improves monotonically over the iterations and converges within approximately $10$ iterations. In addition, a larger number of layers and meta-atoms yields a higher EE at convergence, highlighting the benefit of the increased wave-domain design flexibility offered by larger SIMs. These results demonstrate both the fast convergence and the effectiveness of the proposed algorithm.

\begin{figure}[t!]	
    \centering
    \includegraphics[width=8 cm, height=4 cm]{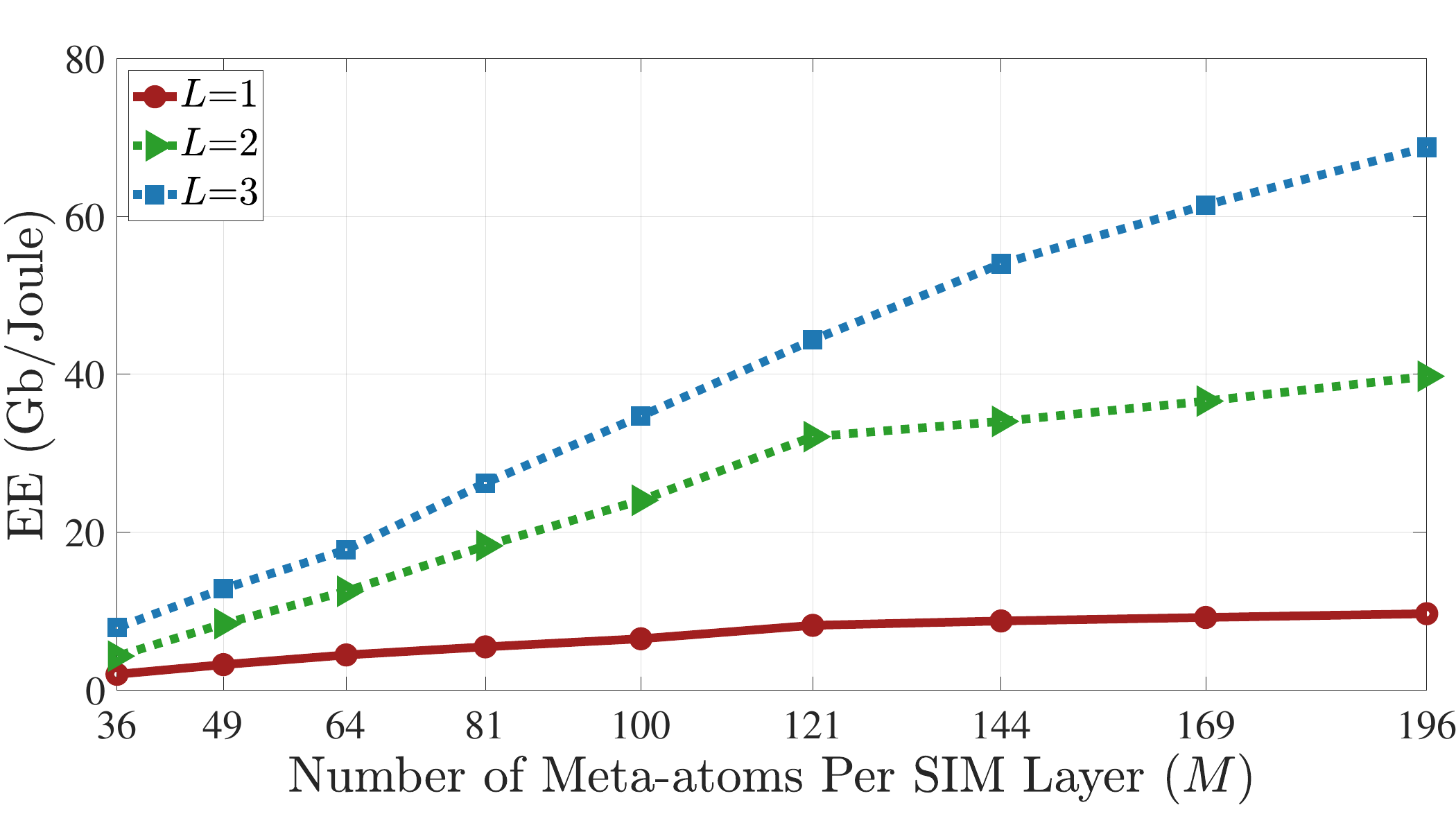}
    \vspace{-5 pt}
    \caption{EE versus the number of meta-atoms per SIM layer.\label{fig:Atoms}}
\end{figure}

\begin{figure}[t!]	
    \centering
    \includegraphics[width=8 cm, height=4 cm]{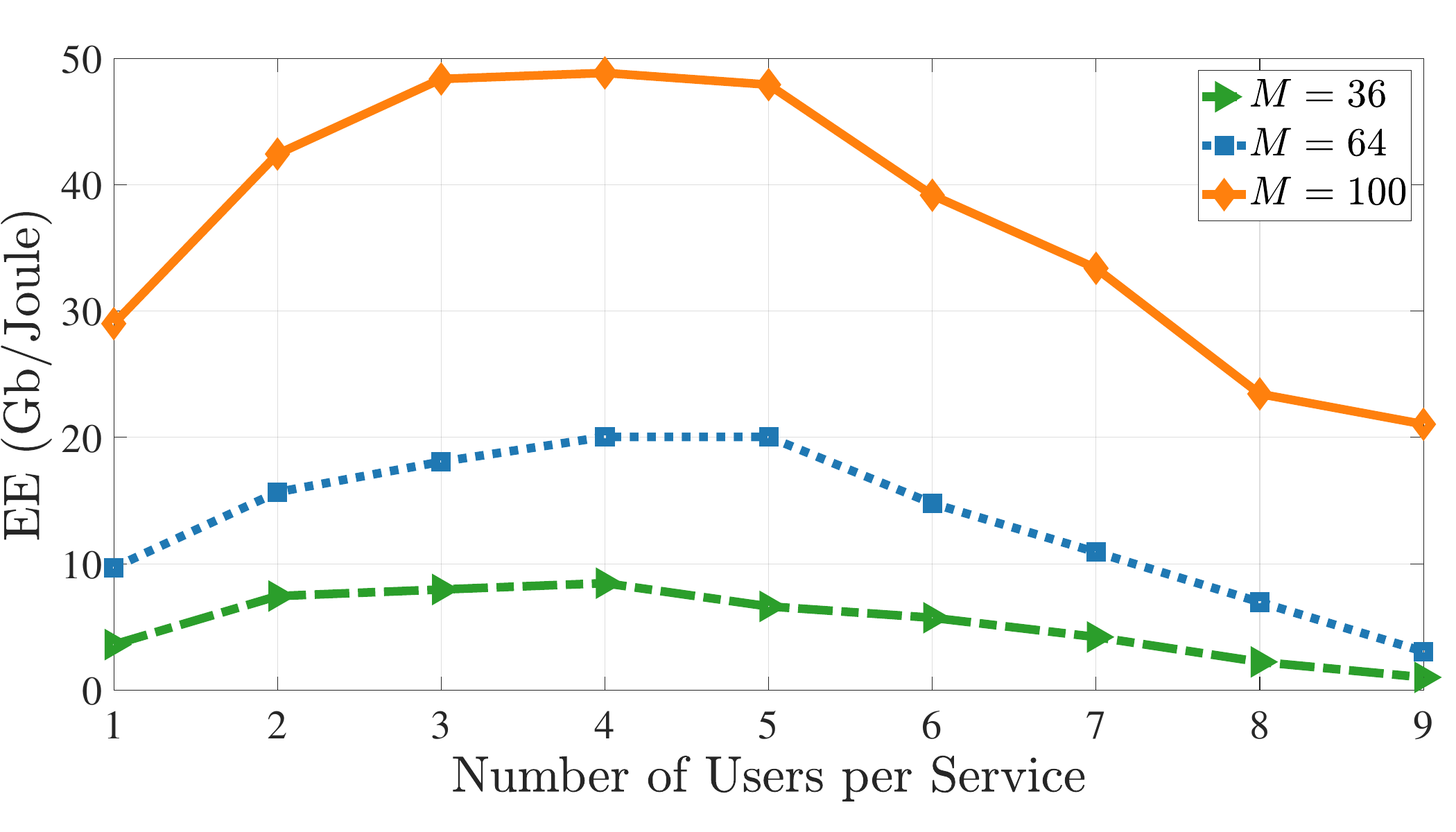}
    \vspace{-5 pt}
    \caption{EE versus the number of users per service.\label{fig:Users}}
\end{figure}

\vspace{-1 em}
\subsection{Impact of SIM Configurations}
In Figs.~\ref{fig:Layers} and \ref{fig:Atoms}, we investigate the impact of the number of SIM layers $L$ and the number of meta-atoms per layer $M$ on the EE, respectively. Here, we use the same parameters as in Fig. \ref{fig:Convergence}. Both $L$ and $M$ can increase the wave-domain degrees of freedom, thereby enhancing wavefront control and improving the EE. Fig.~\ref{fig:Layers} demonstrates that, for a fixed $M$, increasing $L$ initially improves the EE. However, this gain is not monotonic. For smaller $M$ (e.g., $M=16$ and $M=36$), the EE decreases beyond a certain number of layers. The reason is that the penetration loss accumulates with $L$, and when $M$ is relatively small, the resulting beamforming gain is insufficient to compensate for this loss. In contrast, for larger $M$ (e.g., $M=64$ and $M=100$), the increased beamforming flexibility outweighs the penetration loss, and the EE continues to improve with $L$. 
As shown in Fig.~\ref{fig:Atoms}, for a fixed number of layers, increasing $M$ improves the EE. This gain is more pronounced with multiple layers; for example, increasing $M$ from $144$ to $196$ yields about 30\% EE improvement when $L=3$, while the gain reduces to around 10\% when $L=1$. However, the marginal gain diminishes at large $M$, indicating saturation. Therefore, excessively increasing $M$ does not lead to proportional EE gains and may be inefficient in practice.

In Fig. \ref{fig:Users}, we plot the EE versus the number of users per service ($U^{\rm{e}} = U^{\rm{u}}$). For generating this figure, we consider four layers in the SIM. As seen in Fig. \ref{fig:Users}, the EE initially grows as adding users improves the total data rate due to better resource utilization, and the beamforming gain from the SIM enhances the effective channel strength. However, when the number of users exceeds the SIM's spatial degrees of freedom, the ability to perfectly focus beams diminishes. Since the maximum transmit power is limited, this leads to a decrease in EE. Additionally, increasing the number of meta-atoms $M$ in the SIM improves beamforming capability, allowing more users to be served efficiently and shifting the optimal EE point to a higher number of users.

\begin{figure}[t!]	
    \centering
    \includegraphics[width=8 cm, height=4 cm]{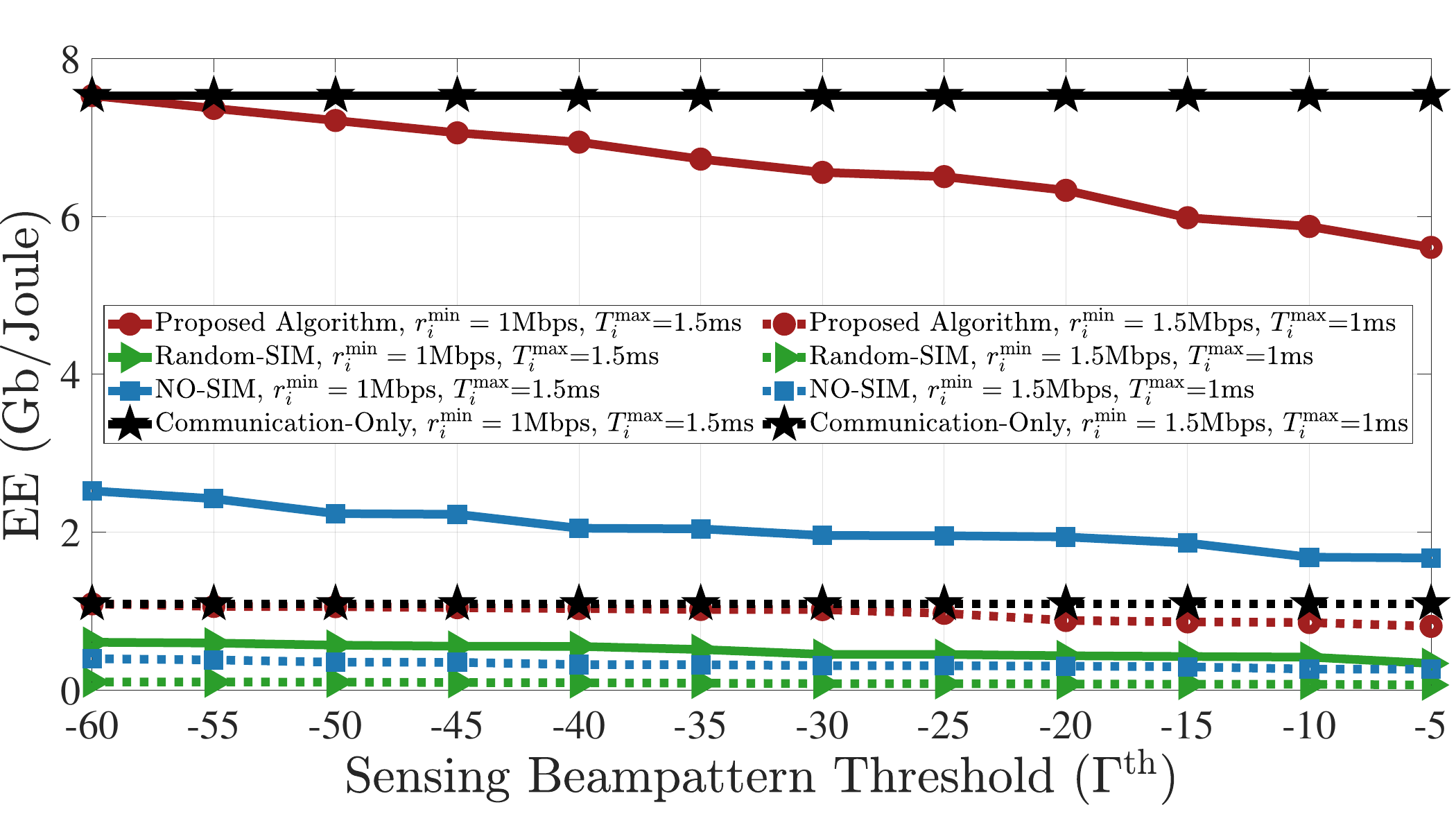}
    \vspace{-5 pt}
    \caption{EE versus the sensing beampattern threshold.\label{fig:Beam}}
\end{figure}
\begin{figure}[t!]	
    \centering
    \includegraphics[width=8 cm, height=4 cm]{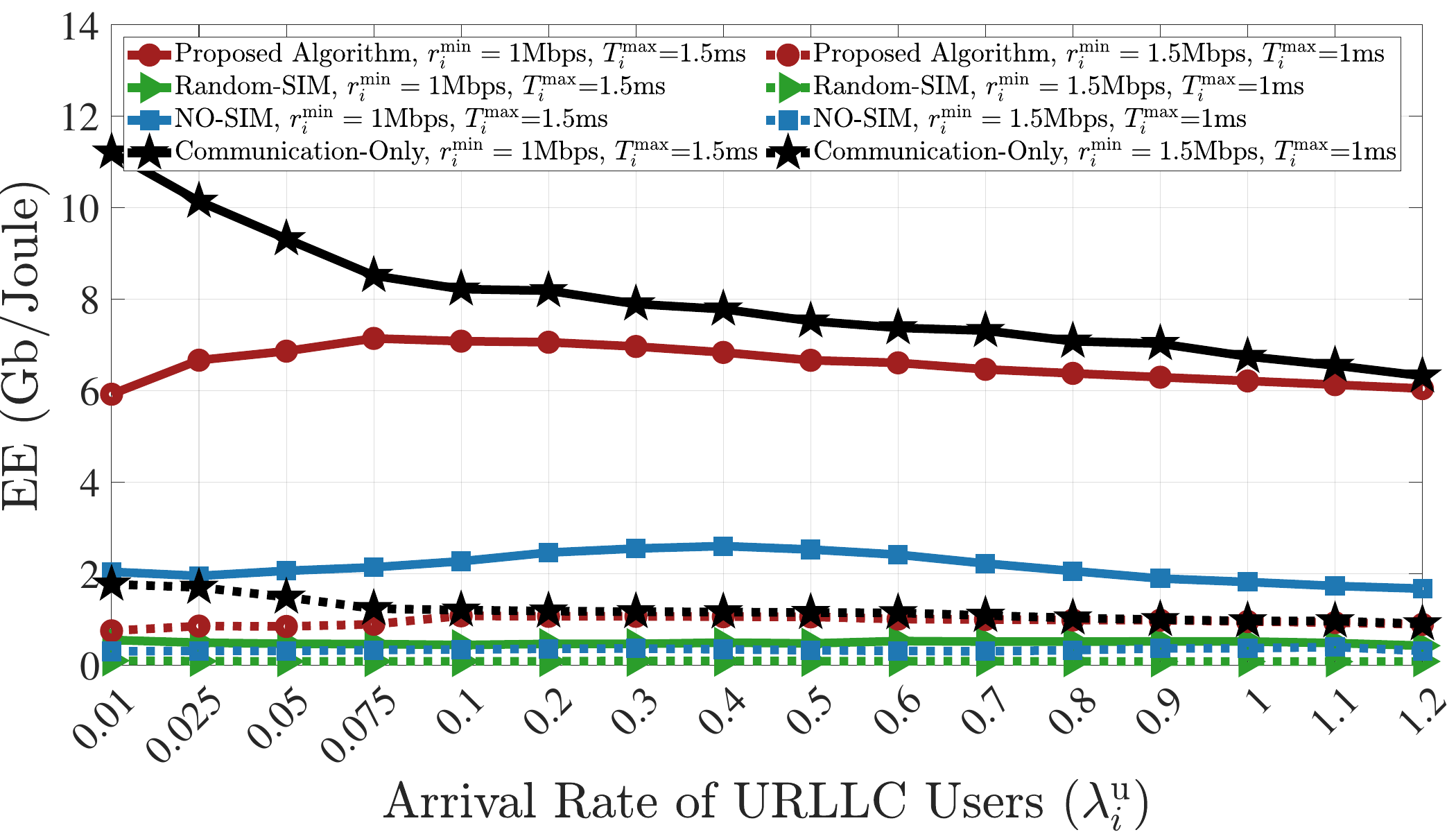}
    \vspace{-5 pt}
    \caption{EE versus the arrival rate of URLLC users.\label{fig:ArrivalRate}} 
\end{figure}

\begin{figure}[t!]	
    \centering
    \includegraphics[width=8 cm, height=4 cm]{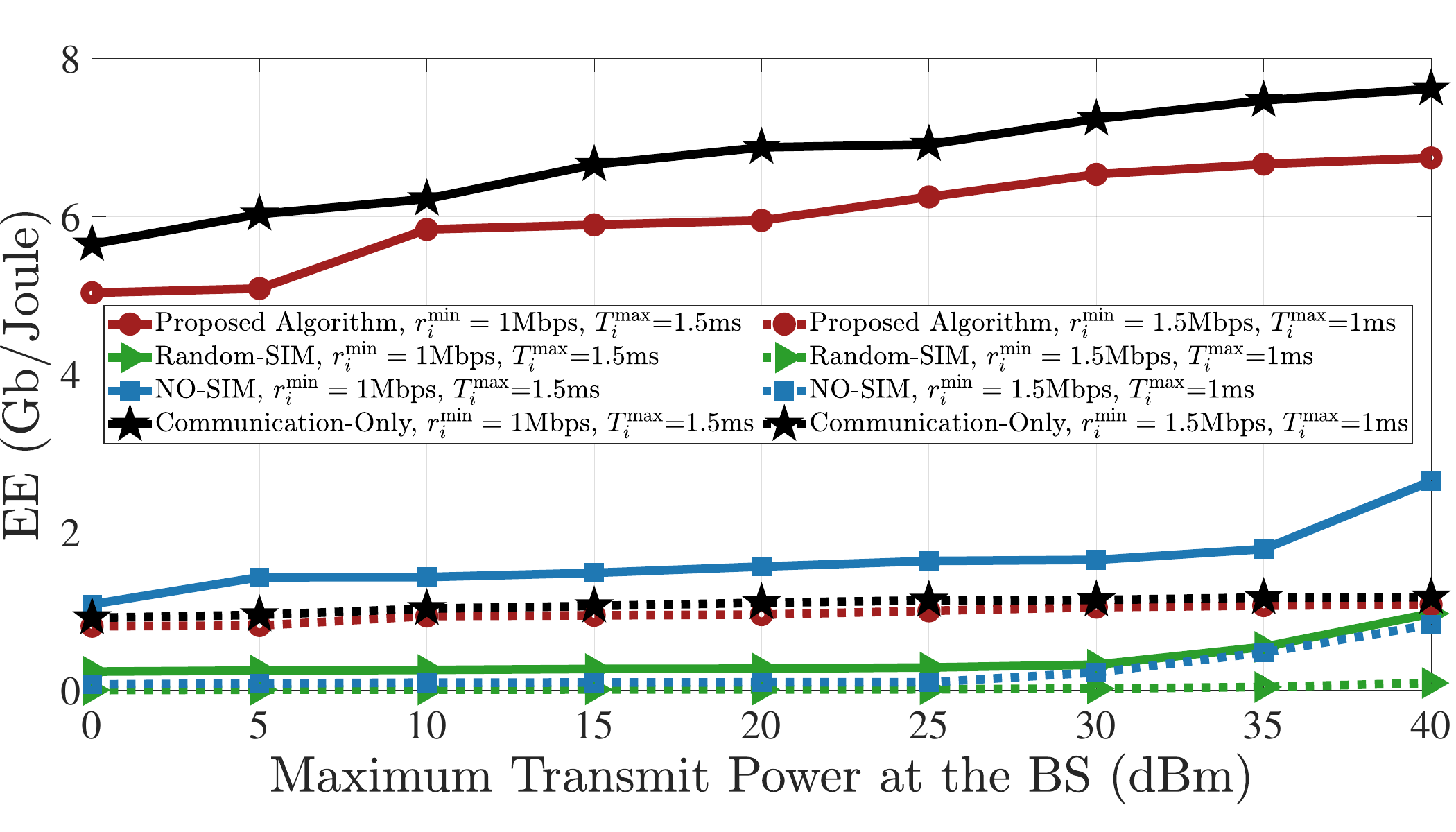}
    \vspace{-5 pt}
    \caption{EE versus maximum transmit power.\label{fig:Power}}
\end{figure}

\begin{figure}[t!]	
    \centering
    \includegraphics[width=8 cm, height=4 cm]{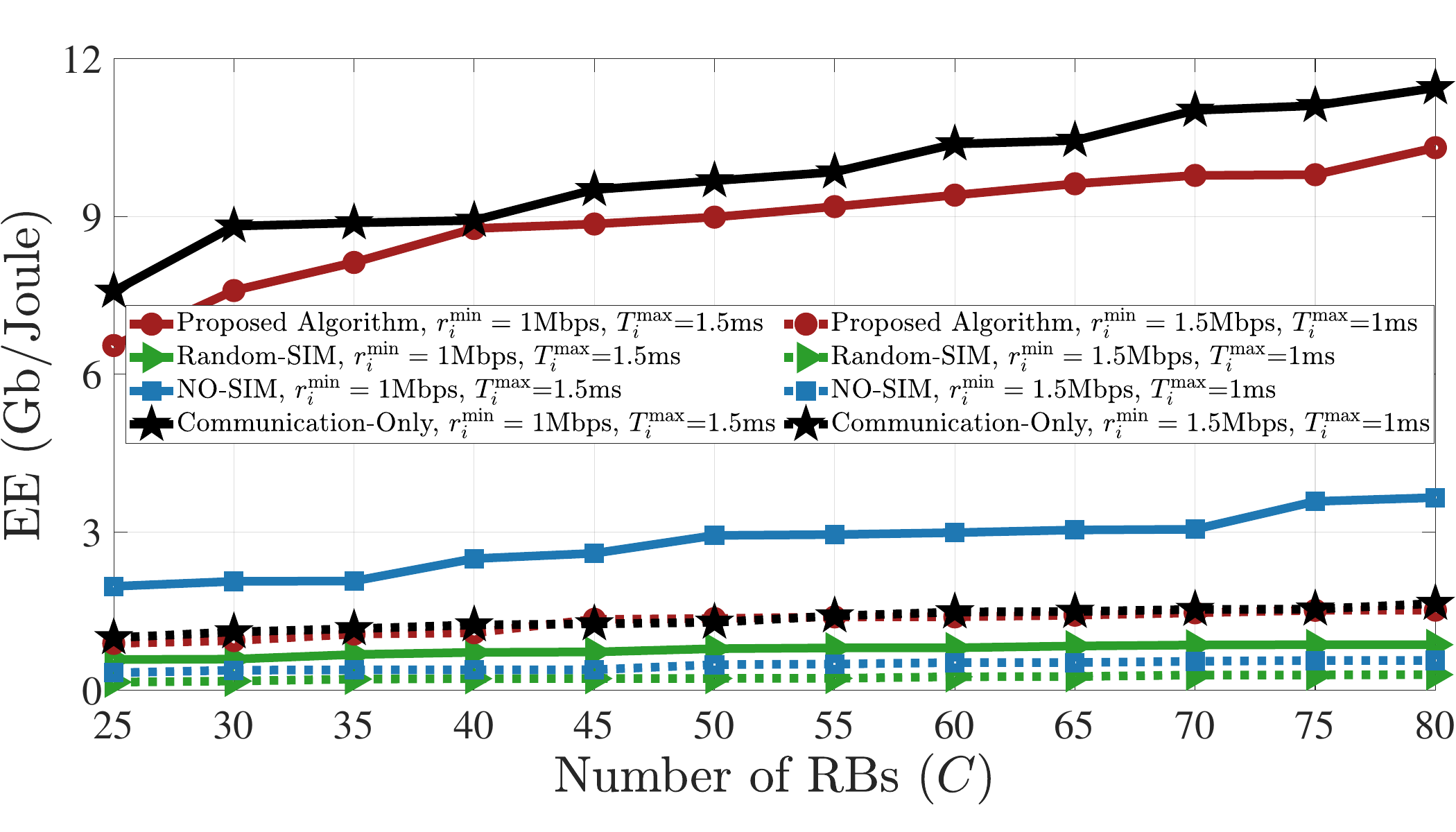}
    \vspace{-5 pt}
    \caption{EE versus the number of RBs.\label{fig:RB}}
\end{figure}
\vspace{-1 em}
\subsection{Impact of Service Requirements}
Next, we investigate the impact of different system parameters on the performance of the proposed algorithm. In addition, to evaluate the effectiveness of the proposed design, we compare it with three benchmark schemes: i) a \textit{Random-SIM} scheme, where the SIM phase shifts are randomly generated without optimization, ii) a \textit{No-SIM} scheme, where the SIM is not deployed, and the system operates without wave-domain control, and iii) a \textit{Communication-Only} scheme, where sensing constraints C4 and C5 are not considered, and the system resources are allocated exclusively to satisfy communication requirements. These baselines allow us to highlight the benefits of jointly optimizing resource allocation and SIM configuration in the proposed algorithm. 

Fig.~\ref{fig:Beam} illustrates the achieved EE versus the sensing beampattern threshold $\Gamma^{\rm{th}}$ for different communication requirements. As $\Gamma^{\rm{th}}$ increases, the sensing constraint becomes more stringent, reducing the degrees of freedom available for communication resource allocation and consequently decreasing the EE across all schemes. Moreover, as the minimum data rate requirement for eMBB users increases and the maximum tolerable latency for URLLC users decreases, higher transmit power is required to meet the corresponding QoS constraints, thereby reducing EE. The proposed algorithm consistently outperforms the \textit{Random-SIM} and \textit{No-SIM} schemes by jointly optimizing communication resources and SIM phase shifts, achieving an EE improvement of 140\% to 230\% over the \textit{No-SIM} scheme. The reason is that the \textit{Random-SIM} scheme cannot fully exploit the wave-domain beamforming capability of the SIM, while the \textit{No-SIM} scheme cannot benefit from the additional beamforming gain enabled by the SIM. Finally, the EE of the communication-only scheme remains unchanged with $\Gamma^{\rm th}$ and provides an upper bound on the EE of ISAC.

The EE versus the arrival rate of URLLC users, $\lambda_{i}^{\rm{u}}$, is shown in Fig.~\ref{fig:ArrivalRate}. It can be seen that the EE of the proposed algorithm first increases and then decreases as $\lambda_{i}^{\rm{u}}$ grows. When the URLLC arrival rate is low, supporting the sensing functionality associated with URLLC users incurs additional power consumption while marginally contributing to the achievable data rate, according to \eqref{EE_URLLC} and \eqref{total_P_URLLC}, resulting in low EE. As $\lambda_{i}^{\rm{u}}$ increases, the joint design of communication and sensing becomes more effective, leading to more efficient resource utilization and thus higher EE. However, when $\lambda_{i}^{\rm{u}}$ becomes large, each URLLC user has more packets to transmit and must satisfy higher rate requirements due to constraint C2 in problem \eqref{Problem_1}, which necessitates increased transmit power and consequently degrades EE. Moreover, the communication-only scheme provides an upper bound on EE, since it does not allocate resources to sensing and thus avoids the additional power consumption associated with sensing functionalities. Nevertheless, the proposed algorithm consistently outperforms the \textit{Random-SIM} and \textit{No-SIM} schemes across all scenarios, demonstrating its effectiveness in balancing communication and sensing functionalities.

\vspace{-1 em}
\subsection{Impact of Network Resources}
\begin{figure}[t!]	
    \centering
    \includegraphics[width=8 cm, height=4 cm]{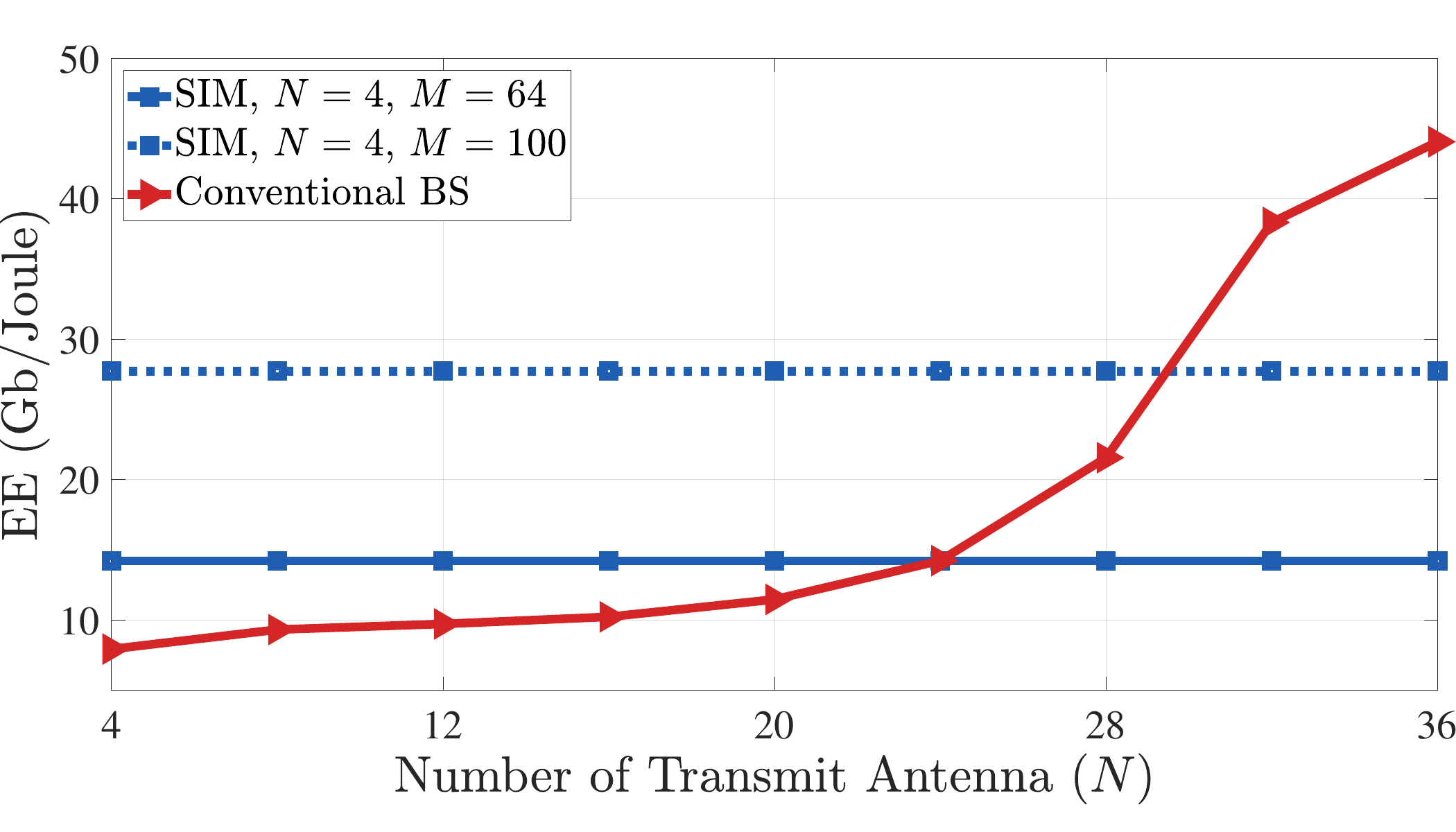}
    \vspace{-5 pt}
    \caption{Comparison of the SIM architecture and the conventional BS scheme versus the number of transmit antennas. \label{fig:Antenna}}
\end{figure}

In Figs. \ref{fig:Power}, \ref{fig:RB}, and \ref{fig:Antenna}, we evaluate the impact of network resource parameters, including the maximum transmit power at the BS ($P^{\rm{max}}$), the number of RBs ($C$), and the number of transmit antennas at the BS ($N$), on the EE, respectively. Fig. \ref{fig:Power} illustrates that as $P^{\rm{max}}$ increases, all schemes exhibit a monotonic improvement in EE due to enhanced power allocation flexibility. Nevertheless, the proposed algorithm demonstrates a more significant performance gain as the transmit power increases, highlighting its superior capability to efficiently exploit the additional power budget and the resulting flexibility in resource allocation. 

In Fig.~\ref{fig:RB}, we depict the EE versus the number of RBs. It can be observed that increasing the number of RBs improves the EE, since a larger RB pool provides higher frequency-domain diversity and more flexibility in RB assignment, thereby enabling more efficient resource utilization. Moreover, the proposed algorithm consistently outperforms the baseline schemes, demonstrating that the joint optimization of communication and sensing variables can more effectively leverage the additional RBs. The communication-only scheme achieves the highest EE, as it operates without sensing constraints C4 and C5. The resulting performance gap highlights the fundamental EE tradeoff introduced by integrating sensing functionality into the communication system.
\begin{figure}[t!]	
    \centering
    \includegraphics[width=8.5 cm, height=4 cm]{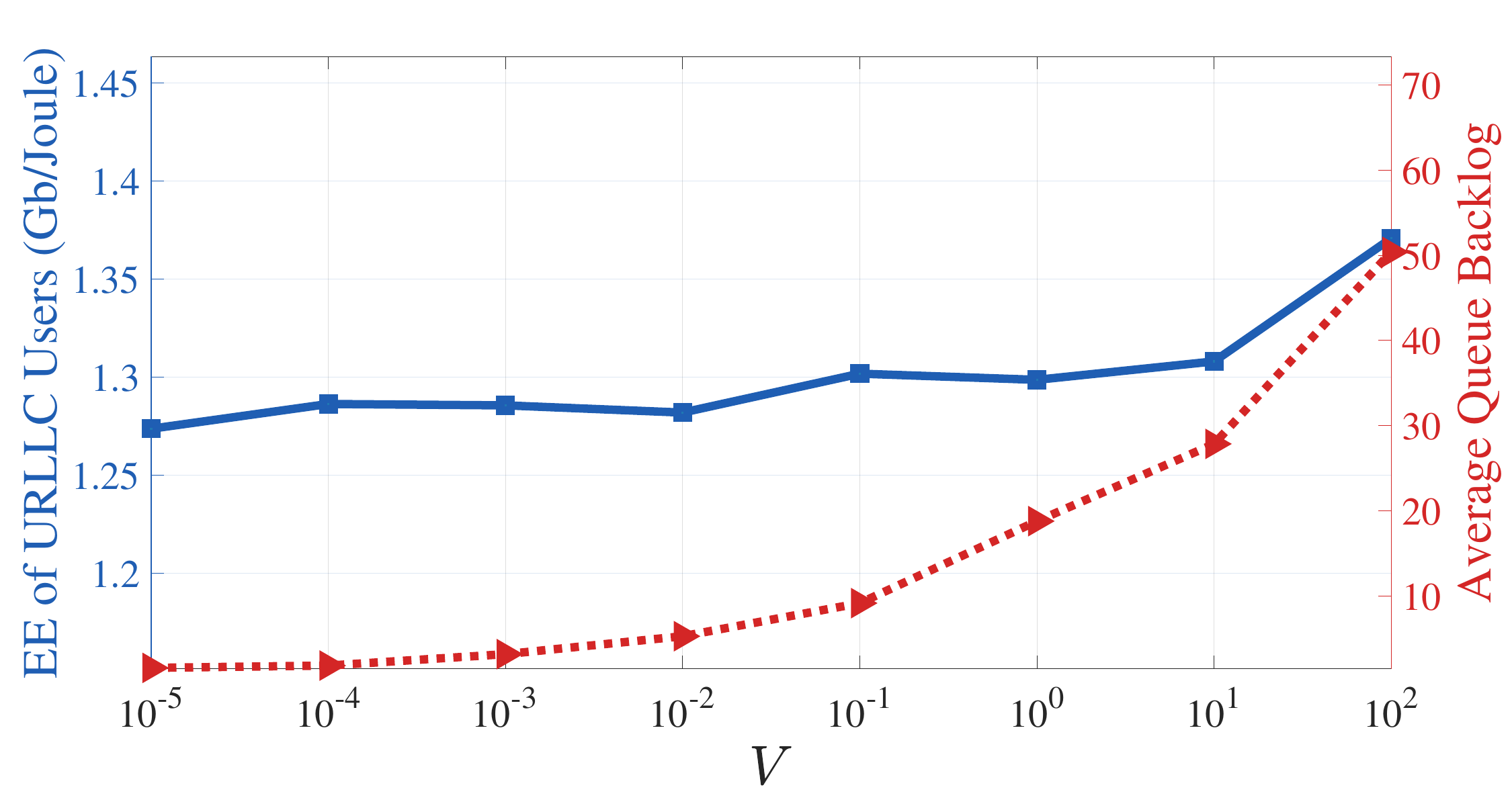}
    \vspace{-5 pt}
    \caption{Trade-off between EE and average backlogs of URLLC users versus $V$.\label{fig:Queue}}
\end{figure}

\begin{figure}[t!]	
    \centering
    \includegraphics[width=8 cm, height=4 cm]{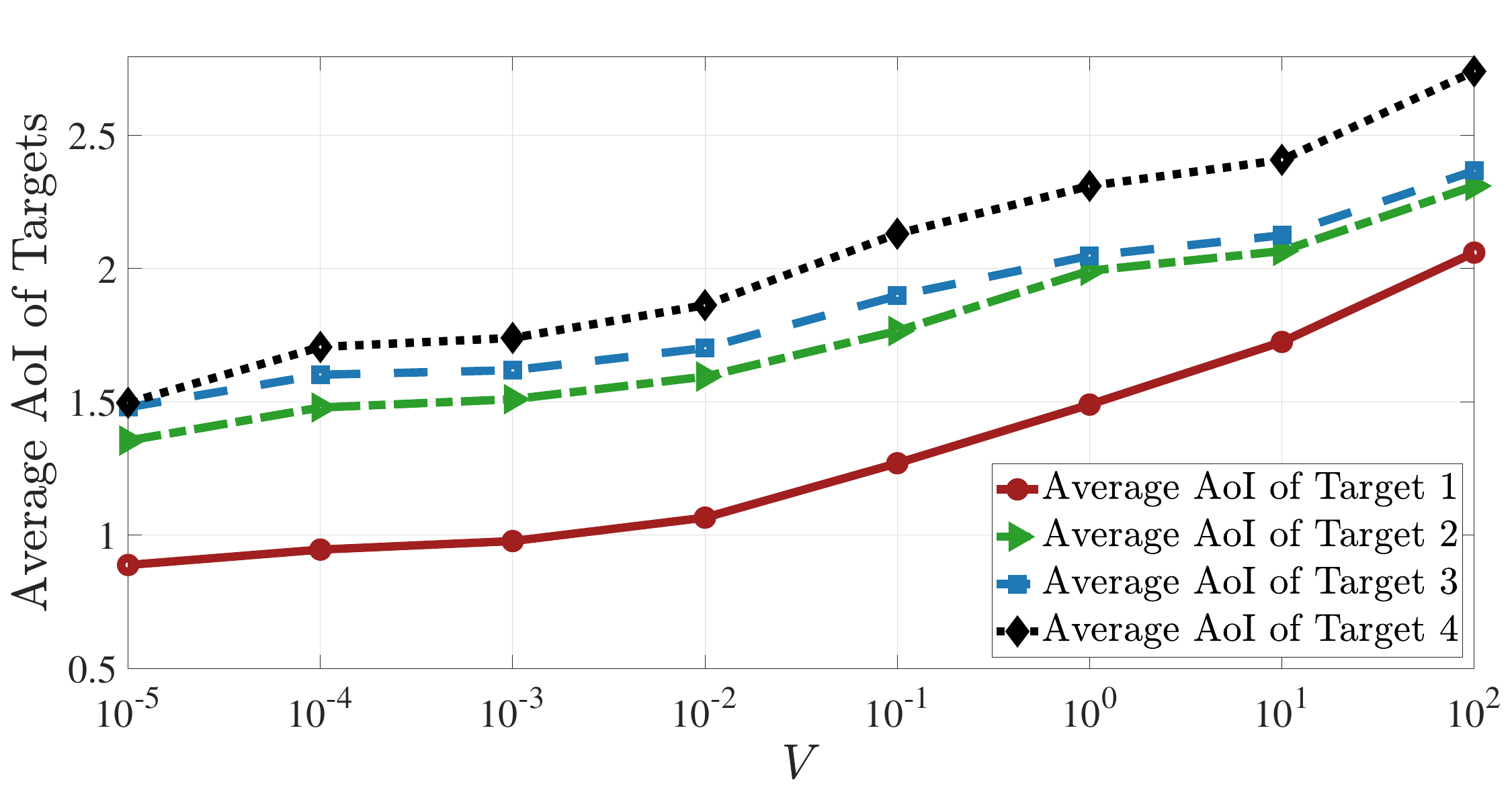}
    \vspace{-5 pt}
    \caption{Average AoI of targets versus $V$.\label{fig:AoI}}
\end{figure}

Fig.~\ref{fig:Antenna} compares the EE of SIM-assisted schemes and the conventional BS as the number of transmit antennas increases from 4 to 36, where the number of SIM layers is $L=3$. It can be observed that the EE of the conventional BS increases with the number of transmit antennas due to the enhanced spatial beamforming gain provided by a larger antenna array. Despite this, when both schemes employ the same number of transmit antennas, i.e., $N=4$, the SIM-assisted system achieves EE improvements of about $79.2\%$ and $147.4\%$ over the conventional BS for $M=64$ and $M=100$, respectively. These gains are enabled by large-aperture wave manipulation and by shifting part of the beamforming from the digital domain at the BS to the SIM. It is worth noting that the conventional BS requires $N=24$ antennas to outperform the SIM-assisted system with $N=4$ antennas and $M=64$ meta-atoms. This performance advantage comes at the cost of increased computational complexity for SIM optimization.

\vspace{-5 pt}
\subsection{Impact of the Lyapunov Control Parameter $V$}
Figs.~\ref{fig:Queue} and \ref{fig:AoI} illustrate the impact of the Lyapunov control parameter $V$ on the communication--sensing trade-off in the considered system. Fig.~\ref{fig:Queue} shows the trade-off between the EE of the communication users and the average backlog of the virtual queues associated with the sensing targets as $V$ varies. As $V$ increases, greater emphasis is placed on the communication users, which leads to an improvement in EE. At the same time, the average backlog of the virtual queues increases. This indicates that the system allocates relatively fewer resources to sensing, which in turn reduces the sensing update frequency. As predicted by the DPP method in Subsection~\ref{DPP}, this behavior reveals the inherent trade-off between optimizing the communication objective and maintaining sensing performance. It is worth noting that the virtual queue backlogs do not correspond to physical queues in the considered system; rather, they are auxiliary variables introduced to enforce the long-term average AoI constraints.

Fig.~\ref{fig:AoI} further illustrates the effect of $V$ on the average AoI of the sensing targets. It can be observed that the average AoI of all targets increases with $V$. This trend is consistent with the growth of the virtual queue backlogs shown in Fig.~\ref{fig:Queue}. Specifically, as $V$ becomes larger, the controller places more emphasis on communication EE and relatively less on sensing updates, thereby decreasing the sensing frequency and increasing the average AoI of the targets. As a result, the AoI constraints become more difficult to satisfy, and for excessively large values of $V$, the threshold $\Delta_i^{\max}$ may be violated. Therefore, $V$ must be properly selected to enhance EE without violating constraint C4.

\section{Conclusion}\label{Con}
We investigated the problem of energy-efficient resource allocation in SIM-aided multi-user ISAC systems under heterogeneous QoS requirements. Specifically, we considered the coexistence of eMBB and URLLC traffic under the puncturing approach, while incorporating sensing requirements in terms of beampattern gain and detection timeliness. To solve the problem, we developed an iterative optimization framework that decomposes it into tractable subproblems, with convex updates for RB allocation and power control and low-complexity updates for SIM phase shifts. Numerical results showed that the proposed scheme achieves EE gains over benchmark schemes while satisfying both communication and sensing requirements. The results also demonstrated the effectiveness of SIM-enabled wave-domain control for ISAC design.
For future work, we aim to extend the proposed framework to multi-cell scenarios with inter-cell interference and imperfect CSI, and investigate AI-driven approaches for real-time resource allocation and SIM-aided ISAC optimization.

%  An EE maximization problem was formulated by jointly optimizing RB allocation, transmit power, and SIM phase shifts, subject to minimum data rate requirements for eMBB users, latency and reliability constraints for URLLC users, and sensing performance constraints.

% \appendix
\balance

\bibliographystyle{IEEEtran}
\bibliography{References}

% \AtNextBibliography{\footnotesize}
% \printbibliography	
% \vspace{-0.7em}	

\end{document}